\documentclass[%
reprint,
nofootinbib,
 amsmath,amssymb,
 aps,
]{revtex4-2}

\usepackage{graphicx}
\usepackage{dcolumn}
\usepackage{bm}
\usepackage[colorlinks=true,citecolor=blue]{hyperref}
\usepackage{cleveref}
\usepackage{subcaption}
\usepackage{listings}

\usepackage{courier}
\begin{document}

\newcommand{\vin}{\textsc{Vincia}}
\newcommand{\py}{\textsc{Pythia}}
\newcommand{\herwig}{\textsc{Herwig}}
\newcommand{\sherpa}{\textsc{Sherpa}}

\newcommand{\pwgbx}{\textsc{Powheg Box}}
\newcommand{\pwgres}{\textsc{Powheg Res}}
\newcommand{\pwgii}{\textsc{Powheg Box v2}}
\newcommand{\pwg}{\textsc{Powheg}}
\newcommand{\rivet}{\textsc{Rivet}}
\newcommand{\fastjet}{\textsc{Fastjet}}

\preprint{MCNET-19-15}

\title{
Coherent Showers in Decays of Coloured Resonances}

\author{Helen Brooks}
 \email{helen.brooks@monash.edu}
\author{Peter Skands}%
\affiliation{%
 School of Physics and Astronomy, Monash University\\
 Wellington Road, Clayton, VIC-3800, Australia
}%


\begin{abstract}
We present a new approach to coherent parton showers in the decays of coloured resonances, based on the notion of ``resonance-final'' (RF) QCD antennae.
A full set of mass- and helicity-dependent $2\to 3$ antenna functions are defined, with the additional requirement of positivity over the respective branching phase spaces.
Their singularity structure is identical to that of initial-final (IF) antennae in $2\to N$ hard processes (once mass terms associated with the incoming legs are allowed for),
but the phase-space factorisations are different. The consequent radiation patterns respect QCD coherence (at leading colour) and reduce to Dokshitzer-Gribov-Lipatov-Altarelli-Parisi and eikonal kernels in
the respective collinear and soft limits. The main novelty in the phase-space factorisation is that branchings in RF antennae impart a collective recoil to the other partons within
the same decay system. An explicit implementation of these ideas, based on the Sudakov veto algorithm, is provided in the \vin~ antenna-shower plug-in to the \py~8 Monte Carlo event generator.
We apply our formalism, matched to next-to-leading order accuracy using \pwg~, to top quark production at the LHC, and investigate implications
for direct measurement of the top quark mass. Finally, we make recommendations for assessing theoretical uncertainties arising from parton showers in this context.

\end{abstract}

\maketitle


\section{\label{sec:intro}Introduction}

In the reconstruction of resonances produced at the Large Hadron Collider,
Shower Monte Carlo (MC) event generators (see \cite{Buckley:2011ms})
play an ongoing critical role. Despite this, many are only formally accurate to 
leading-logarithm, such that there remains a range of ambiguities in their precise 
definition, for example, in the exact form of the splitting kernels used to define  emission
probabilities and Sudakov factors. This in some cases can lead to large theoretical uncertainties in direct
measurements, of which the most notable example is the mass of the top quark.
Nevertheless, analogous to the notion of using ``sensible'' scale choices for evaluating matrix elements of hard processes, 
some ambiguities can be guided by the inclusion of well-motivated physical properties.
One such formally subleading property is that of coherence.

In kinematic limits that correspond to approximately on-shell internal propagators, quantum field theory amplitudes exhibit simple and universal factorisation properties. These  are at the heart of both the treatment of (sequential) resonance decays and bremsstrahlung corrections in high-energy processes. Decay processes in the narrow-width limit, as well as the collinear limits of bremsstrahlung processes, are particularly simple (modulo spin correlations), 
and can be obtained from (squared) Feynman amplitudes which each involve only a single divergent propagator structure. The soft limits, however, characterised by so-called eikonal factors, intrinsically involve a coherent sum over several interfering amplitudes, each with a different  propagator structure. 

\begin{figure}[t]\flushright
\centering
\includegraphics[scale=0.45]{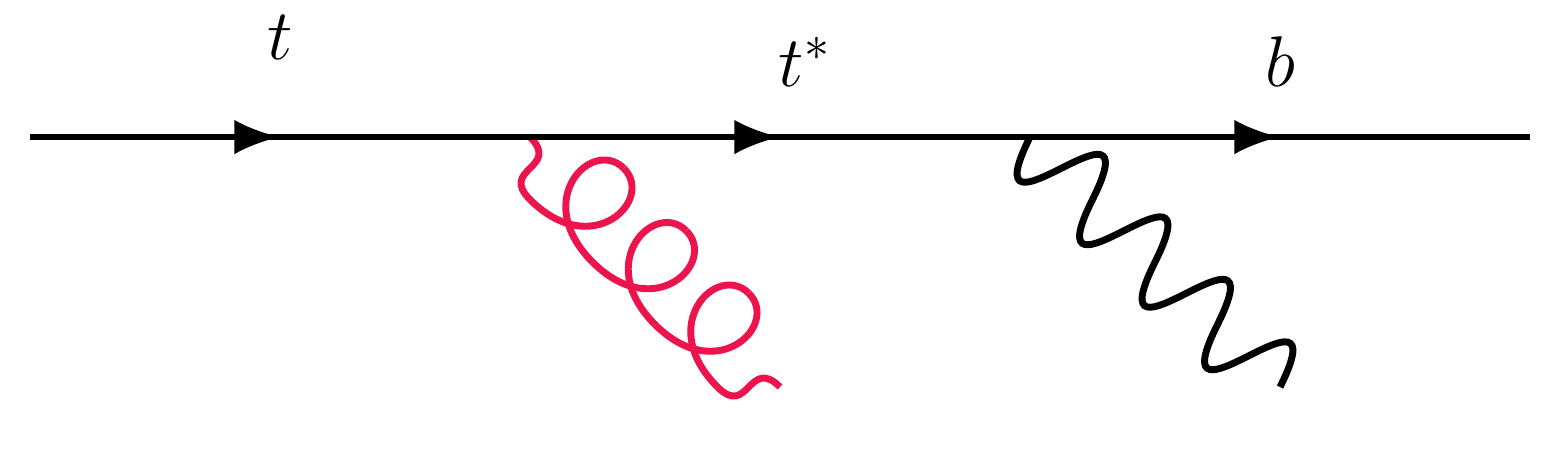}\\[1mm]
\includegraphics[scale=0.45]{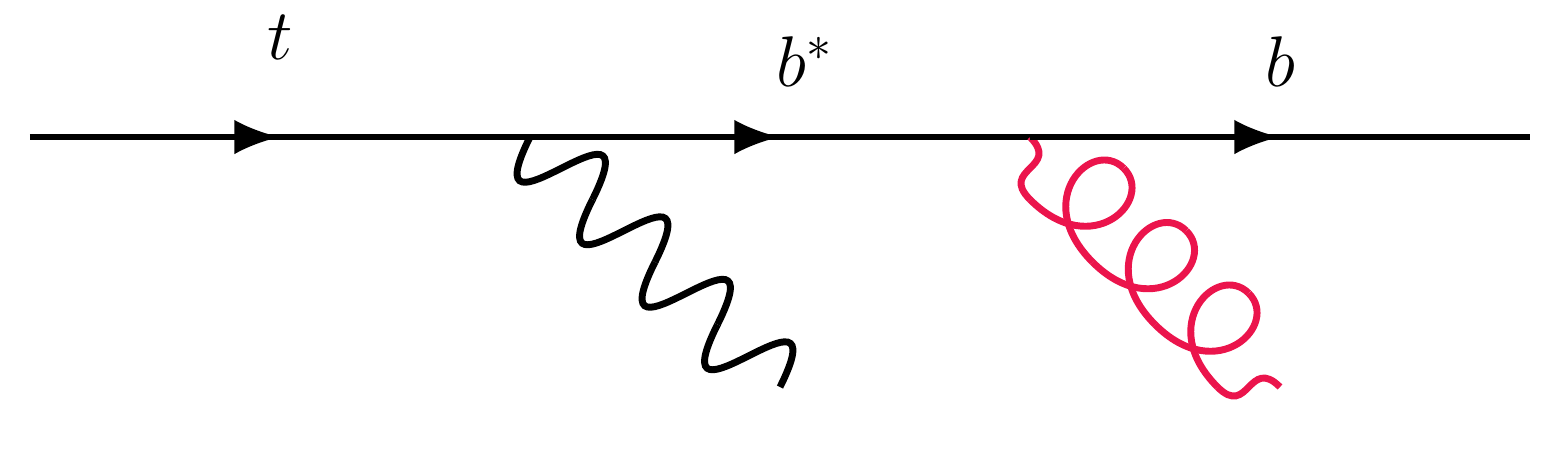}
\caption{The two lowest-order Feynman diagrams that contribute to $t\to bWg$. In both cases, the incoming (outgoing) fermion leg represents an on-shell $t$ ($b$) quark, with mass $m_t$ ($m_b$). In the first diagram, $p^2_{t^*} < m^2_t$. In the second diagram, $p^2_{b^*} > m_b^2$.\label{fig:feynTB}}
\end{figure}

In QCD, one starts from the leading-colour (LC) approximation, which reduces the number of interfering amplitudes that need to be considered to just two for a given gluon becoming soft in a given colour ordering.
These are the two amplitudes that contribute to the corresponding eikonal factor. For the specific case of decays of coloured resonances,
the radiation patterns are normally cast solely in terms of emissions from the produced decay products.
The reasoning for this is that in the rest frame of the decaying particle the contribution to the radiation patterns from the decaying resonance itself can be neglected,
and that the distinction between which particle radiates is anyway gauge dependent and hence unphysical.
Formally, one may partition the full (coherent and gauge invariant) radiation pattern into a term representing radiation from the decaying resonance and one representing radiation from its decay product(s). This is illustrated for top decay in fig.~\ref{fig:feynTB}. 
The former is subdominant (depending on the details of the partitioning, it may even turn out to be negative) and is neglected in most current Shower MC implementations we are aware of. 
It is worth emphasising that matrix-element corrections (MECs) are widely used (e.g., in \py~ and in \pwg) to correct the first emission to the full result;
but in this work we wish to address the issue of coherence in resonance decays more generally, and apply it to all emissions. 

Noting that the antenna-shower formalism (see \cite{Gustafson:1987rq,Giele:2007di,Ritzmann:2012ca}) does not require a partitioning of the radiation pattern into ``radiators'' and ``spectators'',
we derive a set of coherent antenna functions for ``resonance-final'' (RF) colour flows, with full mass- and helicity-dependence. 
We note that these functions exhibit the same singularity structures as corresponding ``initial-final'' (IF) antenna functions derived elsewhere \cite{Ritzmann:2012ca,Fischer:2017htu}, once general mass terms are allowed for in the latter. 
Somewhat arbitrarily we also choose the nonsingular terms to be the same for the IF and RF antenna functions, with minor changes relative to \cite{Ritzmann:2012ca,Fischer:2017htu}  to ensure that all of the IF and RF antenna functions remain positive over all of their respective phase spaces.
This makes them straightforward to interpret in the probabilistic context of a Shower MC. 
We combine these antenna functions with a recoil strategy (alternatively known as a ``kinematics map'') which preserves the 4-momentum of the decaying resonance (and hence in particular its invariant mass),
while imparting a (collective) recoil to the other final-state particle(s) produced in the decay. We argue that this approach should exhibit improved coherence properties over the baseline 
\py~ shower model~\cite{Sjostrand:2004ef,Corke:2010yf}, and that it represents an interesting alternative to other current Shower MC implementations.
We also show that it combines quite naturally with resonance-aware matching in the \pwg~formalism \cite{Nason:2004rx,Frixione:2007vw,Campbell:2014kua,Jezo:2015aia,Jezo:2016ujg}.

Finally, we consider top quark production as a case study for an application of our formalism. This is a particularly well-motivated 
example, since it was recently noted in \cite{Ravasio:2018lzi,FerrarioRavasio:2019glk} that the existing approaches of \py~8.2 \cite{Norrbin:2000uu,Sjostrand:2004ef,Sjostrand:2006za,Sjostrand:2014zea}
and \herwig~7.1 \cite{Gieseke:2003rz,Bahr:2008pv,Bellm:2015jjp,Bellm:2017bvx,Cormier:2018tog} exhibit substantial shape differences in their predictions for 
the differential distribution of the reconstructed invariant mass of the top, already at the level of the parton shower. This has potential implications
for the minimum uncertainty present in the measurement of the top quark pole mass extracted through direct methods. Reducing this uncertainty is desirable since not 
only is the top quark mass an important parameter for many Beyond-the-Standard-Model extensions, but also since the stability of the electroweak vacuum is highly sensitive to its precise 
value \cite{Degrassi:2012ry}.

The outline of the paper is as follows. In \cref{sec:review} we give a review of existing treatments of resonance decays in Shower MCs.
In \cref{sec:showers} we provide details of our new implementation of resonance decays within the \vin\ antenna shower. 
In \cref{sec:matching} we describe resonance-aware matching methods in \pwg, and how these may be used alongside \vin.
In \cref{sec:top} we present our results for top quark production. 
Finally, we summarise in \cref{sec:conclusions}.

\section{Review of Existing Treatments of Resonance Decays}
\label{sec:review}

There are already a range of existing frameworks available for the treatment of resonance decays, so before describing our implementation
we briefly review these alternatives. 

There are a number of components to a parton shower in which there is some flexibility, that must be defined for the shower to be 
fully specified. 
These include the precise form of the splitting kernels in no-emission probabilities, the recoil strategy employed, and the nature of 
the evolution variables used (which determine how ordered sequences of emissions are generated).
One manner of classifying the available options is via the method chosen for organising the singular limits across the set of functions that 
represent the underlying colour-connected objects (each of which is deemed to radiate independently in the leading-colour approximation).
As already noted, in the global antenna-dipole shower framework~\cite{Azimov:1986sf,Gustafson:1987rq,Kosower:1997zr,GehrmannDeRidder:2005cm,Giele:2007di,Ritzmann:2012ca}, a single antenna function contains the entire soft singularity of two colour-connected partons, 
but the collinear limit for gluons is partitioned across two neighbouring antennae.
In the ``partitioned-dipole'' class of showers, both collinear and soft singularities are partitioned across neighbouring dipoles, and in the case of initial-final colour flows, one distinguishes between separate final-initial and initial-final dipole ends (the sum of which is equivalent to a single initial-final antenna in the antenna-shower framework).  
Of this type there are two main variants.

The first is based on Catani-Seymour factorisation, and the form of the splitting kernels are those used in the Catani-Seymour dipole subtraction
method \cite{Catani:1996vz,Catani:2002hc,Schumann:2007mg}.
Such a shower is the default used in the \sherpa~ event generator \cite{Gleisberg:2008ta}, and more recently is also available as an option in 
\herwig~7 \cite{Platzer:2009jq,Bellm:2015jjp,Bellm:2017bvx}.
Here the eikonal is carefully partitioned such that coherence should be recovered after summing over all dipoles.
While this procedure is effective for massless initial state particles, since mass corrections are typically
negative, the initial-final dipole end in resonance decays can become negative. \sherpa~ and \herwig~ offer slightly different solutions to this issue.
In \sherpa~ \cite{Hoche:2014kca}, the resonance is taken to not radiate in decay; instead the entire singularity structure is given to 
the coloured particle in decay, in a manner akin to a sector antenna shower  \cite{LopezVillarejo:2011ap}. All recoil is given to the uncoloured final-state decay product 
(e.g. in $t\rightarrow b W$ the $W$ always takes the recoil.) 
In the \herwig~dipole shower \cite{Cormier:2018tog} the contribution from the initial-final dipole end is neglected entirely; the recoil from the final-initial dipole end is shared out between all the other decay products present (this is a similar approach to our implementation, described in 
\cref{sec:factorisation}). We note that currently the dipole shower option for \herwig~ is only available for strictly on-shell resonances.

Another variant of the partitioned-dipole shower is the transverse-momentum-ordered shower implemented in \py~8 \cite{Sjostrand:2004ef,Sjostrand:2006za,Corke:2010yf,Sjostrand:2014zea}. 
Here, although the individual soft limits for each radiator are reproduced with the vanishing of the ordering variable, it is known that the partition across initial-final dipoles does not
preserve coherence \cite{Skands:2012mm}. 
Despite the lack of coherence, this can be corrected to some extent through matrix-element corrections \cite{Norrbin:2000uu}, which is the default option. 
In addition, ordering in transverse momentum allows for sufficiently compatible definitions such that the multiple-parton interactions (MPI) can
be interleaved with the primary parton shower~\cite{Sjostrand:2004ef,Corke:2010yf}. \py~8 has two options for how recoils are performed in resonance decays. The default option is that after 
the first emission the nearest coloured parton becomes the recoiler. 
Alternatively it is possible to modify this behaviour so that the 
original uncoloured decay product always take the recoil (as in \sherpa).

As representative of the class of partitioned-dipole shower we use \py~8.240~\cite{Sjostrand:2014zea} for later comparisons in \cref{sec:top}, in part because there is already available an interface to recent 
versions of \pwgbx~ \cite{Campbell:2014kua,Jezo:2016ujg,Ravasio:2018lzi}.
In addition since both \vin~ and \py~ share the same  modelling of non-perturbative physics such as hadronisation and underlying event
(although they may differ in the default tuned values of parameters controlling these processes), this better allows us to isolate differences that are of perturbative origin.

An alternative to partitioning the singular limits is for each splitting function to take the full singularity structure, 
and to avoid overcounting via a phase space veto.
In this class coherence is guaranteed, since through the phase space veto each emitter can no longer be considered independent. 
This is the method employed by both sector antenna-showers \cite{LopezVillarejo:2011ap}, in the virtuality-ordered shower in \py~6 \cite{Sjostrand:2006za}, 
and in traditional angular-ordered showers of which the  $\tilde{q}$-shower implemented in 
\herwig~7 is an example \cite{Gieseke:2003rz}. In the latter, the relative opening-angle between colour-connected partons is imposed as an ordering variable, and must
reduce with each subsequent emission. A down-side to angular-ordering is that the phase space factorisation is only approximate, resulting in dead zones
away from the singular limits. The $\tilde{q}$-shower was extended to include resonance decays in \cite{Cormier:2018tog}.
As for \sherpa, the uncoloured decay product is again chosen as the recoiler.
We take the $\tilde{q}$-shower using \herwig~7.1.4 as representative of this class for later comparisons. Again, this is partially motivated by the presence of an existing interface to
\pwgbx~\cite{Ravasio:2018lzi,FerrarioRavasio:2019glk}.

\section{Antenna Showers in Resonance Decays}
\label{sec:showers}

\subsection{Resonance-Final Phase Space Factorisation}
\label{sec:factorisation}

Denoting a generic shower evolution variable by $Q^2$, the no-emission probability for an antenna evolved over the interval $[Q_1^2,Q_2^2]$ is given by the antenna Sudakov factor, $e^{-\mathcal{A}}$, where 
\begin{equation}
 \mathcal{A}(Q_1^2,Q_2^2) = \int_{Q_1^2}^{Q_2^2} \mathrm{d}\Phi_\mathrm{ant} \ 4 \pi \alpha_s \mathcal{C} \bar{a}.
 \label{eq:antIntegral}
\end{equation}
Here $\Phi_\mathrm{ant}$ is the (3-dimensional) $2\to 3$ antenna phase space, $\bar{a}$ is a colour- and coupling-stripped antenna function, and $\mathcal{C}$ is the appropriate colour factor
(for a discussion on the conventions used, see \cite{Giele:2011cb}). 
The antenna function captures the leading singularities of the relevant tree-level matrix elements (but may also contain finite terms in addition).

The antenna phase space depends on a factorisation of the post-branching Lorentz invariant phase space, 
\begin{equation}
\mathrm{d}\Phi_{n+1} = \mathrm{d}\Phi_\mathrm{ant} \times \mathrm{d}\Phi_{n}
\label{eq:antfact}
\end{equation}
in such way that the degrees of freedom of the branching itself and the pre-branching particles can be treated independently.  
Unlike in traditional parton showers where such phase space factorisations only hold in the soft and collinear limits, 
\cref{eq:antfact} is exact. 

We now consider the decay of a coloured resonance $A\rightarrow K + \{X\}$, where $K$ is a final-state particle colour-connected to $A$, and $\{X\}$ schematically denotes any other decay products. 
(For example, in $t\to bW$, the top quark would be identified with $A$, the $b$ quark with $K$, and the $W$ with $X$.)
The phase space measure is simply \cite{GehrmannDeRidder:2009fz}:
\begin{equation}
 \mathrm{d}\Phi_{A\rightarrow  K + \{X\}} = \frac{1}{8(2\pi)^2} \frac{\lambda^{1/2}(m_A^2,m_{AK}^2, m_K^2)}{m_A^2}\mathrm{d}\Omega_K
\end{equation}
where $\lambda(a,b,c) = a^2 +b^2 +c^2 -2ab -2ac -2bc$ is the K\"all\'{e}n function, and $m^2_A = p_A^2$, $m_{AK}^2 = (p_A-p_K)^2 = p_X^2 = m_X^2$ and $m^2_K = p^2_K$. 
There are only two degrees of freedom, representing the global orientation of the frame. 

After a branching from the dipole stretching between $A-K$, we denote the post-branching partons by  $a\rightarrow j k + \{X'\}$, where the prime on $X'$ emphasises that an overall recoil may be imparted to the $X$ system.  
Defining the invariant $s_{jk} \equiv 2 p_j \cdot p_k$ (as opposed to the $m^2_{jk} = (p_j+p_k)^2)$,
the phase space can be written as:
\begin{equation}
 \mathrm{d}\Phi_{a\rightarrow  j k + \{X\}} =  \frac{1}{(4\pi)^5} \frac{\mathrm{d}s_{aj}\mathrm{d}s_{jk} \mathrm{d}\phi}{m_A^2}  \mathrm{d}\Omega_K,
 \label{eq:postbranchingPHSP}
\end{equation}
where $\phi$ corresponds to a rotation of the branching plane about the original orientation of $K$.

The antenna phase space measure is therefore:
\begin{equation}
 \mathrm{d}\Phi_\mathrm{ant}  = \frac{1}{16\pi^2} \frac{\mathrm{d}s_{aj}\mathrm{d}s_{jk}}{\lambda^{1/2}(m_A^2,m_{AK}^2, m_K^2)} \frac{\mathrm{d}\phi}{2\pi}.
 \label{eq:antphsp}
\end{equation}

Implicit in the above derivation is the assumption that the mass of the system of recoilers, $p^2_X = \left(\sum_{i\in\{X\} } p_i \right)^2$ 
is preserved, hence $p_X^2 = p_{X'}^2$, and that this is equivalent to the antenna mass. In addition we impose that the invariant mass of the resonance is unchanged (a feature that is essential for resonance-aware matching),
leading to the identity
\begin{equation}
s_{AK} + s_{jk} + m^2_k + m^2_j - m^2_K = s_{aj} + s_{ak}.
\end{equation}
Finally it is presumed that $j$ and $k$ are produced on-shell.

We now turn to the subject of how the post-branching kinematics are constructed from a given point specified by $s_{aj}$, $s_{jk}$, and $\phi$, subject to
the aforementioned constraints. Such a prescription is called a recoil strategy or \textit{kinematic map}. It is easiest to set up the kinematics in the resonance centre-of-mass frame, such that
\begin{align}
 E_j &= \frac{s_{aj}}{2 m_a},\\
 E_k &= \frac{s_{ak}}{2 m_a},\\
 \cos\theta_{jk} &= \frac{2E_bE_g - s_{jk}}{2 \sqrt{\left(E_k^2 -m_k^2\right)\left( E_j^2 -m_j^2\right)}}.
\end{align}
At this stage there remains an ambiguity regarding rotations $\psi$ in the branching plane (about an axis perpendicular to the dipole axis). 
We specify that $X$ only recoils longitudinally with respect to the dipole axis, and all transverse recoil is shared between $j$ and $k$. 
Finally we rotate by $\phi$ about the dipole axis. 

Following the above construction, we boost back to the lab frame to recover the momentum of the resonance $a$. Each particle in the system $\{X\}$ receives its share of the momentum 
by boosting each by $p_X - p_{X'}$.

We remark that the kinematic map described here is very similar to the prescription recently implemented in \cite{Cormier:2018tog}.

Before concluding this section, we note that had we instead selected a single particle $R\rightarrow r$ to act as a recoiler, it no longer holds that the mass of the antenna is equivalent to the mass
of the recoiler (after the first emission). Supposing we represent the decay as  $A\rightarrow R K +\{X\}$ and $a\rightarrow r + j+ k +\{X\}$ before and after the emission, and by definition
neither $A (=a)$ nor $\{X\}$ recoil, factorisation implies we must preserve $p_R+p_K = p_r + p_j +p_k$. 
Now we also have that
\begin{align}
p^2_R =& (p_A-p_X)^2 +m^2_K -2(p_A-p_X)\cdot p_K, \\
p^2_r =& (p_A-p_X)^2 +m^2_k+m^2_j\nonumber \\
&-2(p_A-p_X)\cdot (p_k+p_j).
\end{align}
Thus it becomes impossible to simultaneously preserve $m_R$ and $m_{AK}$ without violating the factorisation.
It is undesirable to change either; for example in the case of top decays, where a $W$ is selected as the recoiler, the mass should be distributed according to a Breit-Wigner that
is very precisely measured, so it would be inappropriate to give it a large virtuality. On the other hand, sacrificing $m_{AK}$ is tantamount to modifying the factorisation
\cref{eq:antphsp}: everywhere we must replace $A \rightarrow A-X$ and $a \rightarrow a-X$. In addition to modifying the volume of phase space, the identity of the invariants 
is modified with respect to those which appear in the singular part of the real emission matrix elements. Thus a map in which a single particle recoils is 
pathological from the perspective of the antenna formalism. Nevertheless we have implemented such a map for the sake of understanding its effect, and for more equivalent comparisons with
\py.

\subsection{Massive Initial-Final Antenna Functions}
\label{sec:antennae}

The final-final antenna functions used in \vin~ were first derived in~\cite{Giele:2011cb} 
and extended to include mass effects in~\cite{GehrmannDeRidder:2011dm}. 
Massless initial-final and initial-initial antenna functions were presented in \cite{Fischer:2016vfv}.
Finally helicity antennae were added in \cite{Fischer:2017htu} for the massless case.

Here we shall define so-called ``resonance-final'' antennae where both initial- and final-state partons may be massive.
These shall be expressed in terms of the dimensionless invariants, defined as follows:
\begin{align}
 y_{aj} = \frac{s_{aj}}{s_{AK}+s_{jk}}, &&  y_{jk} &= \frac{s_{jk}}{s_{AK}+s_{jk}}, \nonumber \\
 \mu^2_a = \frac{m^2_a}{s_{AK}+s_{jk}}, &&  \mu^2_j &= \frac{m^2_j}{s_{AK}+s_{jk}}, \nonumber \\
 \mu^2_k = \frac{m^2_k}{s_{AK}+s_{jk}}.&&
\end{align}

The mass corrections act to regulate the collinear limit; furthermore they contribute quite large negative corrections
away from the limit. In fact, if only the leading singular terms are retained (for example, the Altarelli-Parisi splitting kernels) these
can become negative. However, by including additional finite terms (that are required to vanish in the soft and collinear limits)
we can guarantee positive-definiteness over the entire physical phase space. 

The full list of helicity-dependent antenna functions may be found in \cref{app:antennae}.
Their singular terms are obtained from the massive helicity-dependent final-state antennae through crossing symmetry,
while their nonsingular terms have been modified to ensure positivity over the full RF and IF branching phase spaces.

The singular parts of the unpolarised antenna functions (defined as the sum of helicity-dependent antennae, averaging over initial helicities)
relevant for top quark decay are, for $q_A q_K\rightarrow q_a g_j q_k$:
\begin{align}
 a^{RF}_{g/qq} = &  
  \frac{1}{s_{AK}}\left[\frac{(1-y_{aj})^2 + (1-y_{jk})^2}{y_{aj} y_{jk}} \right. \nonumber\\
  & \qquad \left. 
  -\frac{2\mu^2_a(1-y_{jk})}{y_{aj}^2} 
  -\frac{2\mu_k^2}{y_{jk}^2}  \right],
  \label{eq:antrf_qqSing}
\end{align}
for $q_A g_K\rightarrow q_a g_j g_k$:
\begin{align}
a^{RF}_{g/qg} = & \frac{1}{s_{AK}} \Bigg[\frac{(1-y_{aj})^3 + (1-y_{jk})^2}{y_{aj} y_{jk}}  \nonumber \\
  & \qquad + (1-\alpha)\frac{1-2y_{aj}}{y_{jk}}- \frac{2\mu^2_a (1-y_{jk}) }{y_{aj}^2}  \Bigg],
\label{eq:antrf_qgSing}
\end{align}
where $\alpha \in [0,1]$ parameterises the partitioning of the collinear singularity of the final-state gluons\footnote{Note that the singular part of the term proportional to $\alpha$ is antisymmetric under interchange of the two final-state gluons, $j\leftrightarrow k$,
hence it cancels when summing over two neighbouring antennae. The default choice is $\alpha = 0$ and may be set with 
\texttt{Vincia:octetPartitioning}}, and for $g\to q\bar{q}$ splittings of the final-state gluon, $q_A g_K\rightarrow q_a q_j \bar{q}_k$:
\begin{equation}
a^{RF}_{q/gX}  = \frac{1}{2m^2_{jk}} \left[ y_{ak}^2 + y_{aj}^2 + \frac{2m_{j}^2}{m_{jk}^2}\right].
\end{equation}

Both of the two emission antennae reduce to the (massive) eikonal in the soft ($y_{jk}\rightarrow 0, y_{aj}\rightarrow 0$) limit:
\begin{equation}
a_\mathrm{eik}  =\frac{1}{s_{AK}}\left[ \frac{y_{ak}}{y_{aj}y_{jk}} - \frac{\mu_a^2}{y^2_{aj}} - \frac{\mu_k^2}{y^2_{jk}}\right].
\end{equation}

These also reproduce the appropriate Altarelli-Parisi splitting functions in the (quasi-)collinear limit \cite{Altarelli:1977zs,Catani:2000ef}:

In addition to being used for coherent branchings in decays of resonances, the same
antennae are used for backwards evolution of the initial state, and are then labelled IF antennae; in this case the initial-state partons are restricted to be massless. This choice is to allow consistency with the
five-flavour massless scheme, since massive initial partons require corrections to the PDFs\footnote{We note, however, that we could in principle use the above antennae also for massive initial state partons, should a set of massive PDFs become available, see e.g.\ recent developments 
in \cite{Krauss:2017wmx,Forte:2019hjc}.}.

As mentioned above, we choose (helicity-dependent) nonsingular terms to ensure positivity of all of the antenna functions over both the RF and IF phase spaces. At the unpolarised level, these sum to:
\begin{align}
 f^{RF}_{g/qq} = &  
\frac{1}{s_{AK}}\left[ 
-\frac{\mu^2_a}{y_{aj}}\Big(
  (1-y_{jk}) 
  + (1-y_{aj})  
  \Big) \right. \nonumber \\
    &\left. + \frac{\mu_k^2}{y_{jk}}\left(  \frac{1}{2}(2-y_{jk})\left(2+\frac{y^2_{aj}}{1-y_{aj}}\right)\right) \right. \nonumber \\
  & \left. + \frac{1}{2}(2-y_{aj})(2-y_{jk}) \right],
  \label{eq:antrf_qqNonSing}
\end{align}
\begin{align}
f^{RF}_{g/qg} = & \frac{1}{s_{AK}} \Bigg[  \frac{\mu^2_a}{y_{aj}} \Big(  (1-y_{aj}) -(2-y_{jk})^2  \Big) \nonumber\\
  & \quad + \frac32 + y_{aj} - \frac{y_{jk}}{2} - \frac{y_{aj}^2}{2} \Bigg],
\label{eq:antrf_qg}
\end{align}
and $f^{RF}_{q/gX} = 0$.

In addition to the above ``resonance-final'' antennae, the following additional initial-final antennae are required to study resonance processes in hadron colliders
(for example $pp\rightarrow t\bar{t}$. For gluon emissions $g_A q_K\rightarrow g_a g_j q_k$ we have:
\begin{align}
 a^{IF}_{g/gq} &=  \frac{1}{s_{AK}}\left[\frac{(1-y_{jk})^3 + (1-y_{aj})^2}{y_{aj}y_{jk}} + \frac{1+y_{jk}^3}{y_{aj}(1-y_{jk})}\right. \nonumber \\
  &\left. - \frac{2\mu_k^2}{y_{jk}^2}\left(1 - \frac{y_{jk}}{4}(3-3y^2_{jk}+y^3_{jk})\left(2+\frac{y^2_{aj}}{1-y_{aj}}\right)\right) \right. \nonumber \\
  &\left. +\frac{1}{2}(2-y_{aj})(3-y_{jk}+y^2_{jk}) \right].
\end{align}
and for $g_A - g_K\rightarrow g_a  g_j g_k$ we have:
\begin{align}
 a^{IF}_{g/gg} =  \frac{1}{s_{AK}} &\left[ \frac{(1-y_{aj})^3 + (1-y_{jk})^3}{y_{aj}y_{jk}} \right. \nonumber \\
 & \left. + \frac{1 + y_{jk}^3}{y_{aj} (1-y_{jk})} \right. \nonumber \\
 &\left.+ (1-\alpha)\frac{1-2 y_{aj}}{y_{jk}} + 3 -2y_{jk}\right].
\end{align}

All other antennae may be found in \cite{Fischer:2016vfv,Fischer:2017htu}.


\subsection{Evolution variables}

\begin{figure*}
\centering
 \begin{subfigure}{0.45\textwidth}
 \centering
 \includegraphics[width=\textwidth]{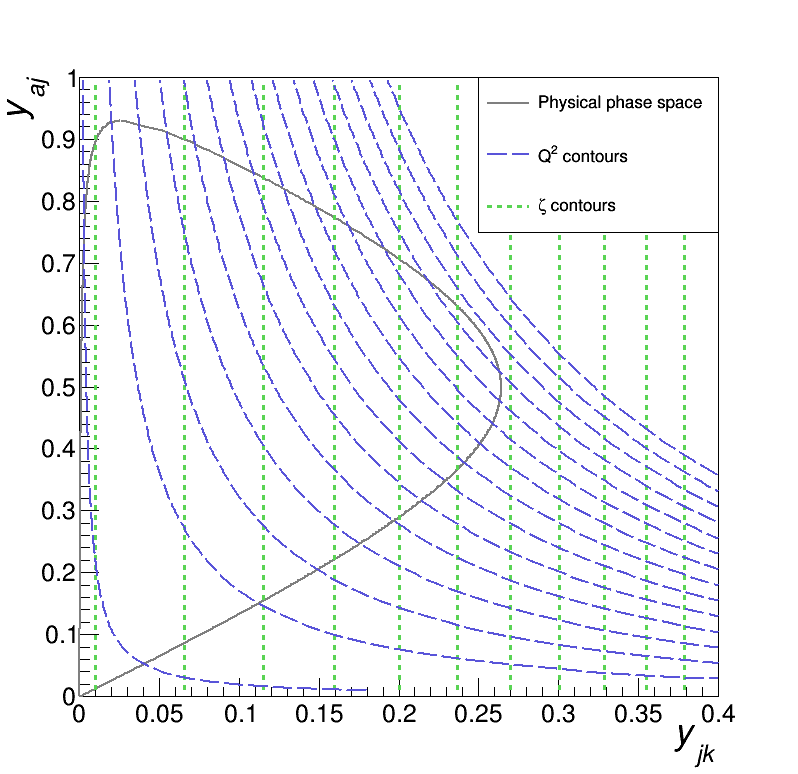}{}
  \caption{}
  \label{fig:emitphasespace}
 \end{subfigure}
  \begin{subfigure}{0.45\textwidth}
 \centering
 \includegraphics[width=\textwidth]{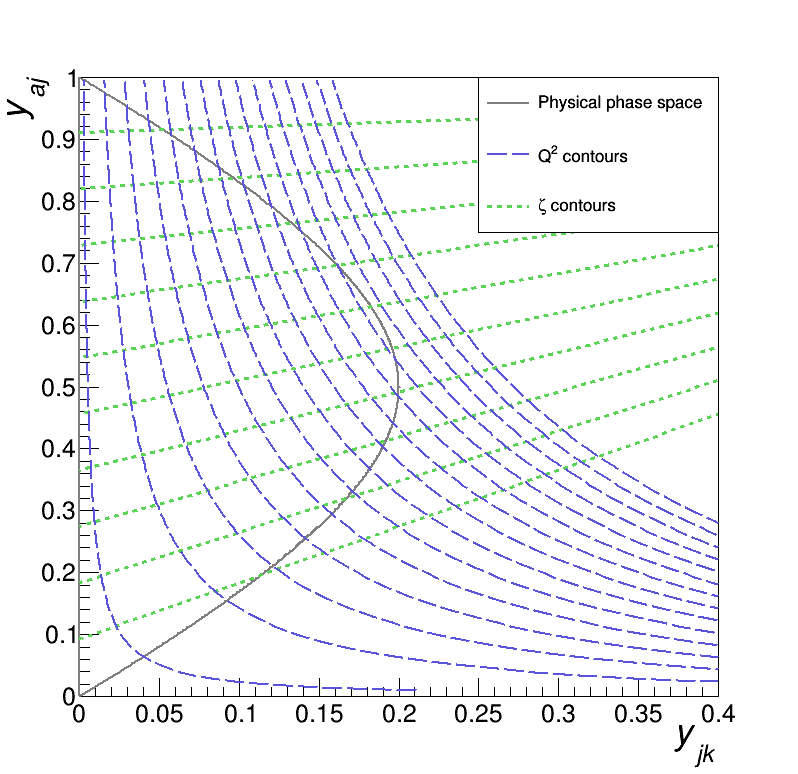}{}
  \caption{}
  \label{fig:splitphasespace}
 \end{subfigure}
\caption{Figure showing contours of equally spaced constant $Q^2_\mathrm{evol}$ (long dashes) and $\zeta$ (short dashes) in the $y_{jk}$, $y_{aj}$ plane
for the case of \subref{fig:emitphasespace} resonance emissions and \subref{fig:splitphasespace} resonance splittings. In the former we took
$m_A=m_a$=171 GeV, $m_K=m_k$=4.8 GeV, $m_X$=80.4 GeV, $m_j$=0 GeV. For the latter we took 
$m_A=m_a$=171 GeV, $m_K$=0 GeV, $m_X = 0.6 m_A$ , $m_j = m_k$=4.8 GeV.
The physical phase space is delineated by the solid grey line.}
\label{fig:phasespace}
\end{figure*} 

Having demonstrated the desired factorisation we may construct an 
ordering variable $Q^2_\mathrm{evol}$ and complementary splitting variable
$\zeta$ through a change of variables from $s_{aj}$, $s_{jk}$.
It is worth remarking that there is relative freedom in the choice for these
variables: all that is required of $Q^2_\mathrm{evol}$ is that it vanishes in the 
soft and collinear limits. The choice for $\zeta$ must be linearly independent of $Q^2_\mathrm{evol}$
and curves of constant $\zeta$ should intersect those of constant $Q^2_\mathrm{evol}$ once and only once.
While different choices of $Q^2_\mathrm{evol}$ should all produce the same result in the soft-collinear limits, 
they will give rise to subleading differences, which in some cases can be quite significant. 
For a more in-depth discussion on this point we refer the reader to \cite{Giele:2011cb}. Since we take the Jacobian associated with the transformation from ($s_{aj},s_{jk}$) to ($Q^2_\mathrm{evol},\zeta$) explicitly into account, the choice of $\zeta$ variable only affects the efficiency of the phase-space sampling and not the final physical distributions.

In the context of interleaved showers \cite{Sjostrand:2006za} it is desirable for the different shower components (e.g., RF and FF antennae for resonance decays, and II, IF, and FF antennae for hard processes) to use similar ordering variables, so that the common sequence of decreasing values of the ordering variables is physically meaningful as a globally decreasing resolution scale. After the first emission in the decay of a resonance, emissions from the RF antenna will compete with emissions from FF antennae in the same decay system. Our choice must therefore be consistent with \textsc{Vincia}'s $p_T$-ordering variable for FF antennae, which is the same as that used in \textsc{Ariadne}~\cite{Gustafson:1987rq,Lonnblad:1992tz},
\begin{equation}
(p_{Tj}^{FF})^2 = 
  \frac{s_{ij} s_{jk}}{s_{ijk}}~.
\end{equation}

For the case of gluon emissions we take:
\begin{equation}
 Q^2_\mathrm{evol,emit} = \frac{s_{aj}s_{jk}}{s_{AK}+s_{jk}},
\end{equation}
while for gluon splittings (to quarks with mass $m_q$) we have:
\begin{equation}
 Q^2_\mathrm{evol,split} = \frac{(s_{aj}-m_q^2)(s_{jk}+ 2 m_q^2)}{s_{AK}+s_{jk}+ 2 m_q^2}.
\end{equation}

There is no requirement upon the choices for $\zeta$ to be equivalent; therefore convenient
choices are selected that are simple, and that allow for the definition of a separable 
trial integral (as we discuss later in \cref{sec:trialintegral}).
We therefore choose:
\begin{equation}
 \zeta_\mathrm{evol,emit} = \frac{s_{jk}+s_{AK}}{s_{AK}}
\end{equation}
for emissions, and
\begin{equation}
 \zeta_\mathrm{evol,split} = \frac{s_{ak}}{s_{AK}}
\end{equation}
for splittings.

In \cref{fig:phasespace} we plot contours of constant of $Q^2_\mathrm{evol}$ and $\zeta$
in the $y_{jk}$, $y_{aj}$ plane.

\subsection{Trial Integral}
\label{sec:trialintegral}

\begin{figure}
\vspace{1em}
\centering
 \begin{subfigure}{0.33\textwidth}
 \centering
 \includegraphics[width=\textwidth]{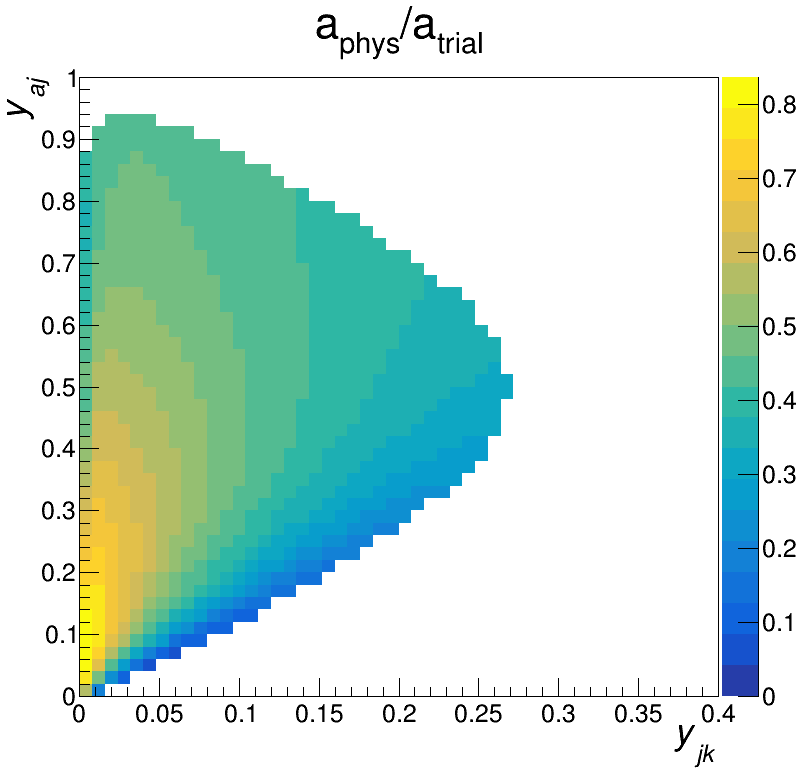}{}
  \caption{}
  \label{fig:antratioemit}
 \end{subfigure}
  \begin{subfigure}{0.33\textwidth}
 \centering
 \includegraphics[width=\textwidth]{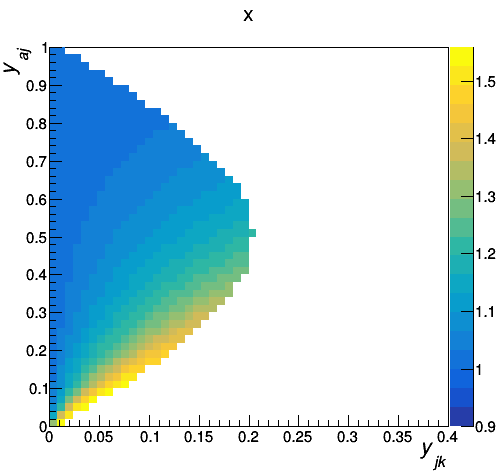}{}
  \caption{}
  \label{fig:normplit}
 \end{subfigure}
  \begin{subfigure}{0.33\textwidth}
 \centering
 \includegraphics[width=\textwidth]{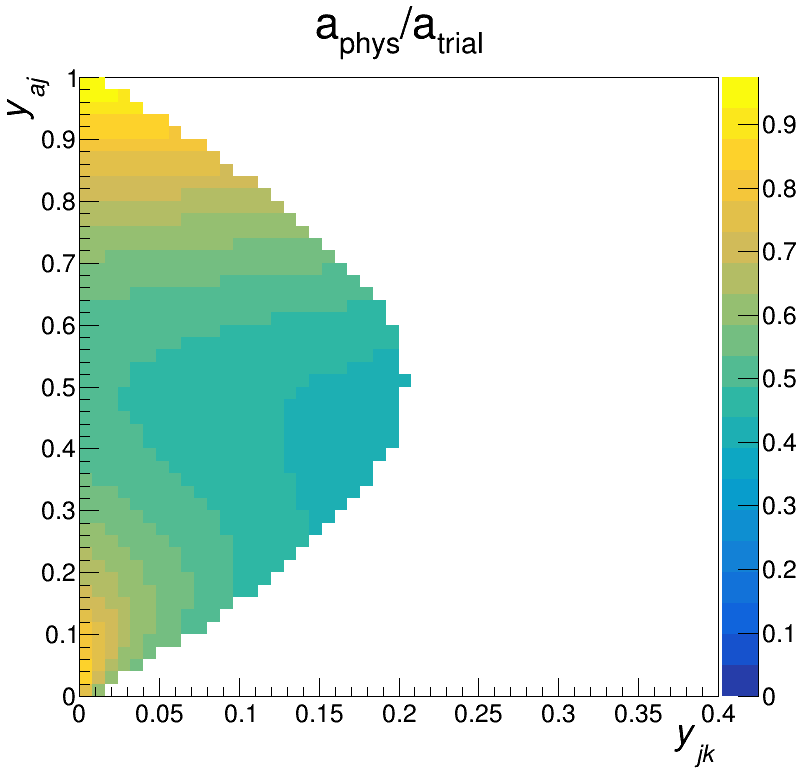}{}
  \caption{}
  \label{fig:antratiosplit}
 \end{subfigure}
\caption{Heat map showing the ratio of the physical unpolarised antennae for resonance-final branchings (as given in \cref{sec:antennae,app:antennae})
to the trial antennae given in \cref{sec:trialintegral} for (\subref{fig:antratioemit}) emissions and (\subref{fig:antratiosplit}) splittings.
The peculiar shape of contours for the latter is in part due to the multiplicative factor $x$ in \cref{eq:trialsplit}, which we 
plot in \subref{fig:normplit}. The masses used to generate these plots are as for \cref{fig:phasespace}.}
\label{fig:antratio}
\end{figure} 

In the context of the Sudakov veto method \cite{Sjostrand:2006za}, we generate trial branchings 
by solving
\begin{equation}
 r = e^{-\mathcal{A}_\mathrm{trial}(Q^2_\mathrm{max},Q^2)}
 \label{eq:trialgen}
\end{equation}
for $Q^2$ given some random number $r \in(0,1)$. Here $\mathcal{A}_\mathrm{trial}$
is the trial integral, obtained by evaluating \cref{eq:antIntegral} for some trial
antenna function $a_\mathrm{trial}$, over a phase space volume equal to or greater than
the physical phase space.

The trial antennae must be an overestimate to the physical antenna functions
given in \cref{sec:antennae} at every point in phase space; namely they must capture the
leading singular behaviour.
They must also be simple enough such that both \cref{eq:trialgen} and its inverse are 
analytically calculable.

Starting with emissions, the change of variables is given by
\begin{equation}
 \mathrm{d}s_{aj}\mathrm{d}s_{jk} = s_{AK} \frac{\zeta}{\zeta-1} \mathrm{d}Q^2_\mathrm{evol}\mathrm{d}\zeta
\end{equation}
For the trial antenna integral, we take:
\begin{equation}
 a_\mathrm{trial,emit} = 2\frac{s_{AK}+s_{jk}}{s_{aj}s_{jk}} = \frac{2}{Q^2_\mathrm{evol}}
\end{equation}
This choice captures the leading double soft singularity; in \cref{fig:antratioemit} 
we demonstrate numerically that it is a suitable overestimate everywhere in phase space.

Putting everything together we get the following expression for the trial integral:
\begin{align}
 \mathcal{A}_\mathrm{trial}(Q^2_\mathrm{max},Q^2) =& \frac{2\mathcal{C}s_{AK} (I(\zeta_\mathrm{max}) - I(\zeta_\mathrm{min}))}{\lambda^{1/2}(m_A^2,m_{AK}^2, m_K^2)} \nonumber \\
&\cdot \int^{Q^2_\mathrm{max}}_{Q^2} \frac{\mathrm{d}\tilde{Q}^2}{\tilde{Q}^2}\frac{\alpha_s(\tilde{Q}^2)}{4\pi}.
\end{align}
where we note that we have averaged over the azimuthal angle $\phi$, and the integral over $\zeta$ is given by:
\begin{equation}
 I(\zeta) = \ln\left((\zeta-1)e^{\zeta -1}\right).
\end{equation}
The trial integral for $Q^2$ depends upon whether fixed or one-loop running of $\alpha_s$
is used\footnote{Even if two-loop running of $\alpha_s$ is desired, one-loop running is 
performed for the 
trial integral using the 2-loop value of $\Lambda_\mathrm{QCD}$; this overestimates the two-loop running result and is corrected by including the ratio of $\alpha_s$ in the accept probability.}, but in either case this is straightforward to perform and invert. Having generated a trial
$Q^2$ we  generate $\zeta$ by inverting
\begin{equation}
 r =  \frac{I(\zeta) - I(\zeta_\mathrm{min})}{I(\zeta_\mathrm{max}) - I(\zeta_\mathrm{min})},
\end{equation}
to give
\begin{equation}
 \zeta(r) = 1 + W\left[e^{ ( I(\zeta_\mathrm{max}) - I(\zeta_\mathrm{min})r +I(\zeta_\mathrm{min})} \right],
\end{equation}
where $W(z)$ is the Lambert W function that is the inverse to $ze^z$ (which we implemented according to the method in \cite{Veberic}).

Moving onto $g\to q\bar{q}$ splittings, the change of variables is now:
\begin{equation}
 \mathrm{d}s_{aj}\mathrm{d}s_{jk} = s_{AK} \frac{s_{AK}+s_{jk}+2m^2_q}{s_{aj}-m^2_q}\frac{1}{x}\mathrm{d}Q^2_\mathrm{evol}\mathrm{d}\zeta,
\end{equation}
where $x$ is a dimensionless factor given by
\begin{equation}
 x = 1 + \frac{(s_{jk}+2m^2_q)(s_{ak}+m^2_q)}{(s_{AK}+s_{jk}+2m^2_q)(s_{aj}-m^2_q)},
\end{equation}
which exhibits the property of being everywhere positive, and tends to unity in the collinear limit (as may be seen in \cref{fig:normplit}).
A suitable choice for the trial antenna is therefore
\begin{equation}
 a_\mathrm{trial,split} =\frac{x}{2(s_{jk}+2m^2_q)}
 \label{eq:trialsplit}
\end{equation}
which captures the leading collinear singularity and exceeds the physical antenna function everywhere
in phase space. This is demonstrated numerically in \cref{fig:antratiosplit}.

The trial integral is now given by
\begin{align}
 \mathcal{A}_\mathrm{trial}(Q^2_\mathrm{max},Q^2) =& \frac{\mathcal{C}s_{AK} (\zeta_\mathrm{max} - \zeta_\mathrm{min})}{2\lambda^{1/2}(m_A^2,m_{AK}^2, m_K^2)} \nonumber \\
&\cdot \int^{Q^2_\mathrm{max}}_{Q^2} \frac{\mathrm{d}\tilde{Q}^2}{\tilde{Q}^2}\frac{\alpha_s(\tilde{Q}^2)}{4\pi},
\end{align}
and $\zeta$ is sampled flatly:
\begin{equation}
 \zeta = (\zeta_\mathrm{max}- \zeta_\mathrm{min})r +\zeta_\mathrm{min}.
\end{equation}


\section{Resonance-Aware Matching with POWHEG}
\label{sec:matching}

In order to seriously assess theoretical uncertainties, for example in the context of direct top mass measurements at the
LHC~\cite{Aad:2015nba,Aaboud:2016igd,Aaboud:2017mae,Aaboud:2018zbu,Chatrchyan:2013boa,Sirunyan:2017idq,Sirunyan:2018gqx,Sirunyan:2018goh,Sirunyan:2018mlv}, 
it is clearly desirable to attain the highest combined logarithmic and fixed-order accuracy that is currently available for the production of
resonances. 

The matching of parton showers to next-to-leading order accuracy through both subtractive (e.g. MC@NLO) \cite{Frixione:2002ik} and 
multiplicative (e.g. \pwg) \cite{Nason:2004rx,Frixione:2007vw} methods
for massive final states has been available for some time \cite{Frixione:2003ei,Frixione:2007nw,Alioli:2010xd}.
For an accurate reconstruction of resonances however, it is important to correct not only the hardest emission in production, but also in the decay
of the resonance. It was noted in \cite{Campbell:2014kua} that a naive application of the \pwg~ method to decays (for example as attempted in \cite{Garzelli:2014dka}) 
in which the kinematics map between the real-emission $\Phi_r$ and the Born $\Phi_B$ phase spaces that modified the virtuality of the resonance, would certainly fail.
This conclusion comes from the realisation that differences in the virtuality of the resonance between 
the $\Phi_r$ and $\Phi_B$ kinematics that exceed its width, spoil the cancellation between the 
the virtual and real-emission contributions to the cross section. Thus the reweighting of the Born cross section could become arbitrarily large, leading to considerable
distortion of the resonance peak. Such difficulties were also expected to be present for subtractive matching methods, as these too could modify the invariant mass
of the resonance. 

To resolve these issues, a so-called ``resonance-aware'' matching method  was proposed and implemented in \pwgii~ \cite{Campbell:2014kua}.
This method performs the next-to-leading-order (NLO) calculation in the narrow width approximation, but applies finite-width effects in an approximate way. 
Essentially this involves generating fully off-shell resonances for the Born phase space, and mapping to an on-shell phase space to perform the
\pwg~ generation of real emissions, before finally mapping onto the real emission phase space to recover the original resonances' virtualities. 
The events are then reweighted to reproduce the correct off-shell next-to-leading-order cross section. 
In addition this method also includes spin correlations to NLO accuracy. 
As alluded to earlier, an essential requirement for consistency with the NLO calculation is that the parton shower must preserve the invariant mass of the resonance. 

Recently the resonance-aware matching method was extended to include exact width effects \cite{Jezo:2015aia}, before being automated in the generator \pwgres~ 
and applied to the process $pp\rightarrow b\bar{b}\ell^+\ell^-\nu_{\ell} \nu_{\bar{\ell}}$ (\texttt{bb4l}) in \cite{Jezo:2016ujg}. In this method, one must specify the resonance from which a given emission originates. 
This is straightforward provided there is a single resonance chain, but is not normally possible where there are interfering resonant diagrams. 
Nevertheless such topologies must be considered in order to extend to exact width effects, as they are technically necessary
to preserve gauge invariance. Thus a modification was required to perform the assignment of an emission to a given resonance. 
The solution was the selection of a given ``resonance history'' based on a partition of the singular regions of phase space. 

In \cite{Ravasio:2018lzi} a comparison of the two resonance-aware matching methods and the much earlier \texttt{hvq} implementation 
for observables relevant to the top mass measurement was performed, using interfaces to \py~8.2 and the angular-ordered $\tilde{q}$-shower in \herwig~7.1.
It was generally observed that the differences between these two showers for a given method much exceeded the differences resulting from 
the different choices for the matching method. In the following section where we consider similar observables, we therefore choose to use
\pwgii~ rather than the more recent generator, as the former is notably faster and we do not expect interference effects to affect our conclusions. 
We use the setting
\begin{lstlisting}
allrad = 1 
\end{lstlisting}
such that emissions are \textit{always} generated in both the decay and production of the resonance. (This is in addition to the default setting,
\begin{lstlisting}
nlowhich = 0
\end{lstlisting}
which controls whether or not emissions are generated in decay at all.)

Where comparisons to \py~8.2 are \herwig~7.1 performed we closely follow the prescription in \cite{Ravasio:2018lzi}. 
For matching to \vin, the procedure we use is very similar to that used for \py.
It is necessary to ensure that \vin~ does not perform an emission harder than the scale of the emission generated by \pwg. 
For emissions in production this information is provided via the \texttt{scalup} value in the Les Houches event file, so this must be 
used as the starting scale in the shower. While for \py~ this behaviour is activated through the settings
\begin{lstlisting}
TimeShower:pTmaxMatch = 1
SpaceShower:pTmaxMatch = 1
\end{lstlisting}
in \vin, the corresponding setting is :
\begin{lstlisting}
Vincia:QmaxMatch = 1
\end{lstlisting}

In addition, the \texttt{UserHooks} class provided as part of the \texttt{bb4l} package, that is employed for the vetoing of radiation 
in resonance decays in \py, (this being flexible enough also for use with \texttt{ttbardec} in \pwgii) was modified for use with \vin.
The algorithm therein is essentially the same, except for minor modifications for interpreting the event record.

\section{Applications to Top Quark Physics}
\label{sec:top}

\subsection{Validation of kinematic map}
\label{sec:validate_kinmap}

\begin{figure}[h]
\centering
 \begin{subfigure}{0.4\textwidth}
 \centering
 \includegraphics[width=\textwidth]{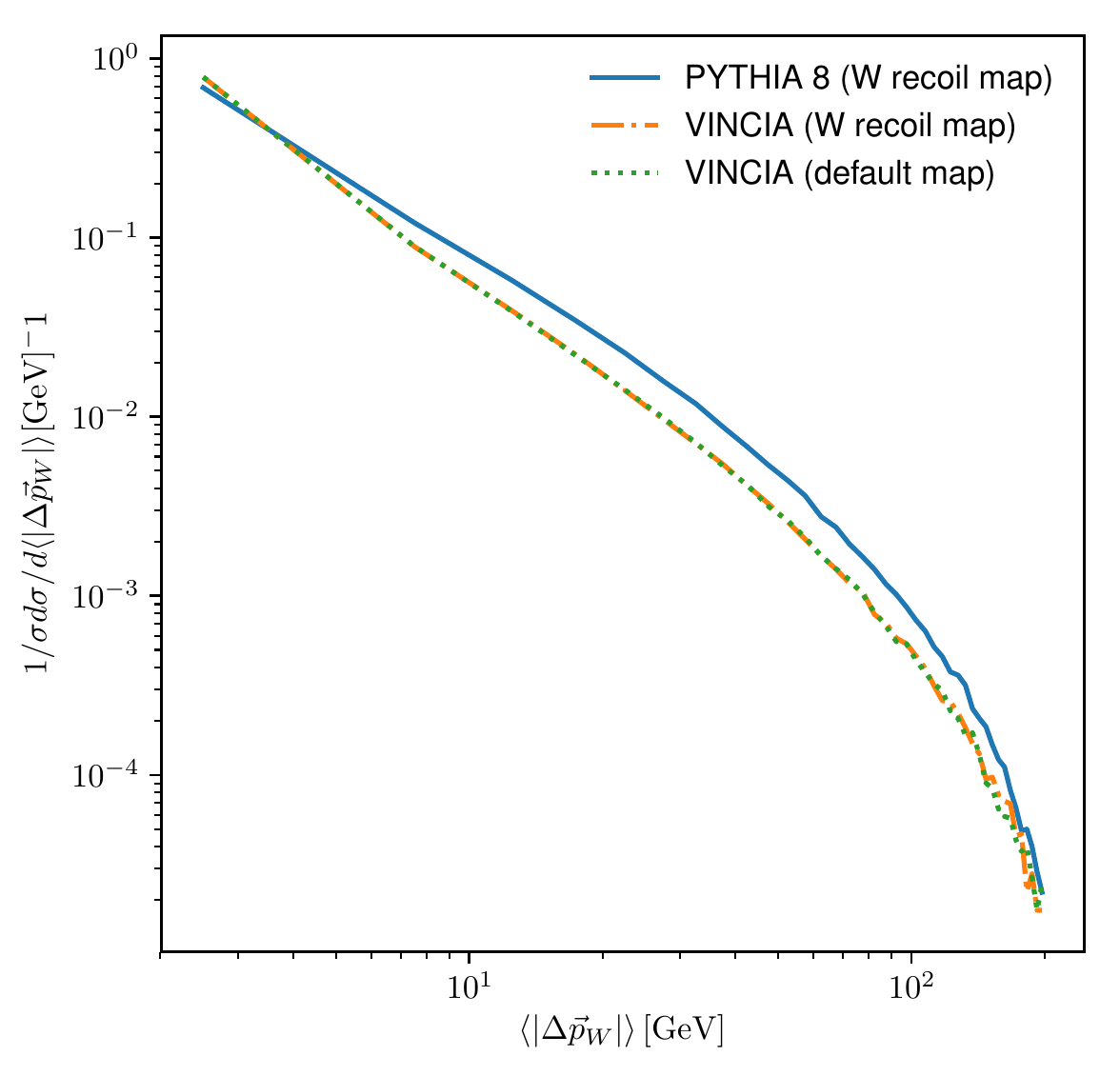}{}
  \caption{}
  \label{fig:recoilafterone}
 \end{subfigure}
 \begin{subfigure}{0.4\textwidth}
 \centering
 \includegraphics[width=\textwidth]{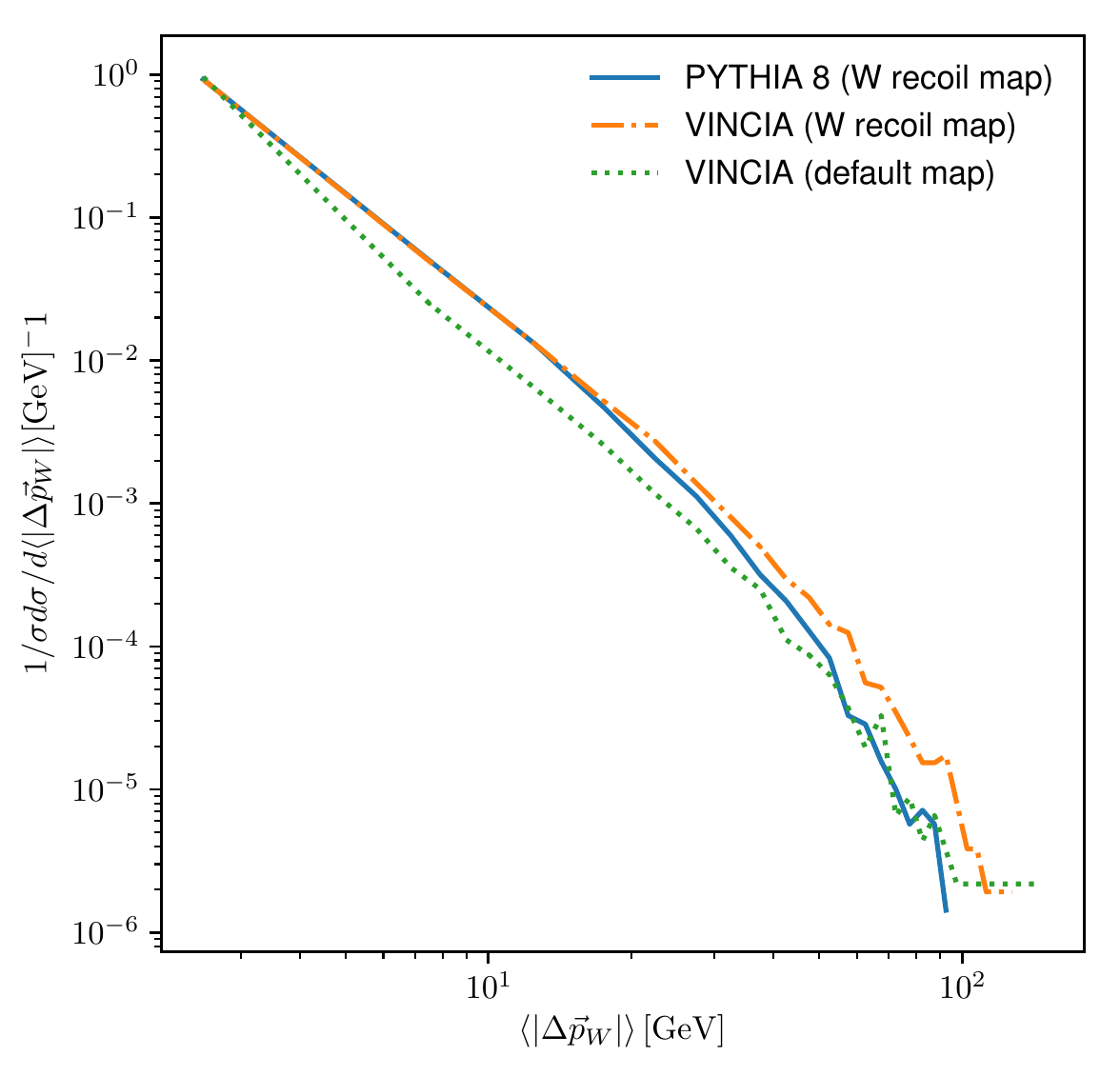}{}
  \caption{}
 \label{fig:recoilaftertwo}
 \end{subfigure}
 \vspace{1em}
\caption{Plots showing the distribution in the change in the three-momentum of the $W$ boson before and after the first emission
\subref{fig:recoilafterone}, and between the first and second emission \subref{fig:recoilaftertwo}, from the resonance-final antenna in $t\rightarrow bWX$ decays.}
\label{fig:wrecdist}
\end{figure}

\begin{figure}[h]
\centering
 \begin{subfigure}{0.4\textwidth}
 \centering
 \includegraphics[width=\textwidth]{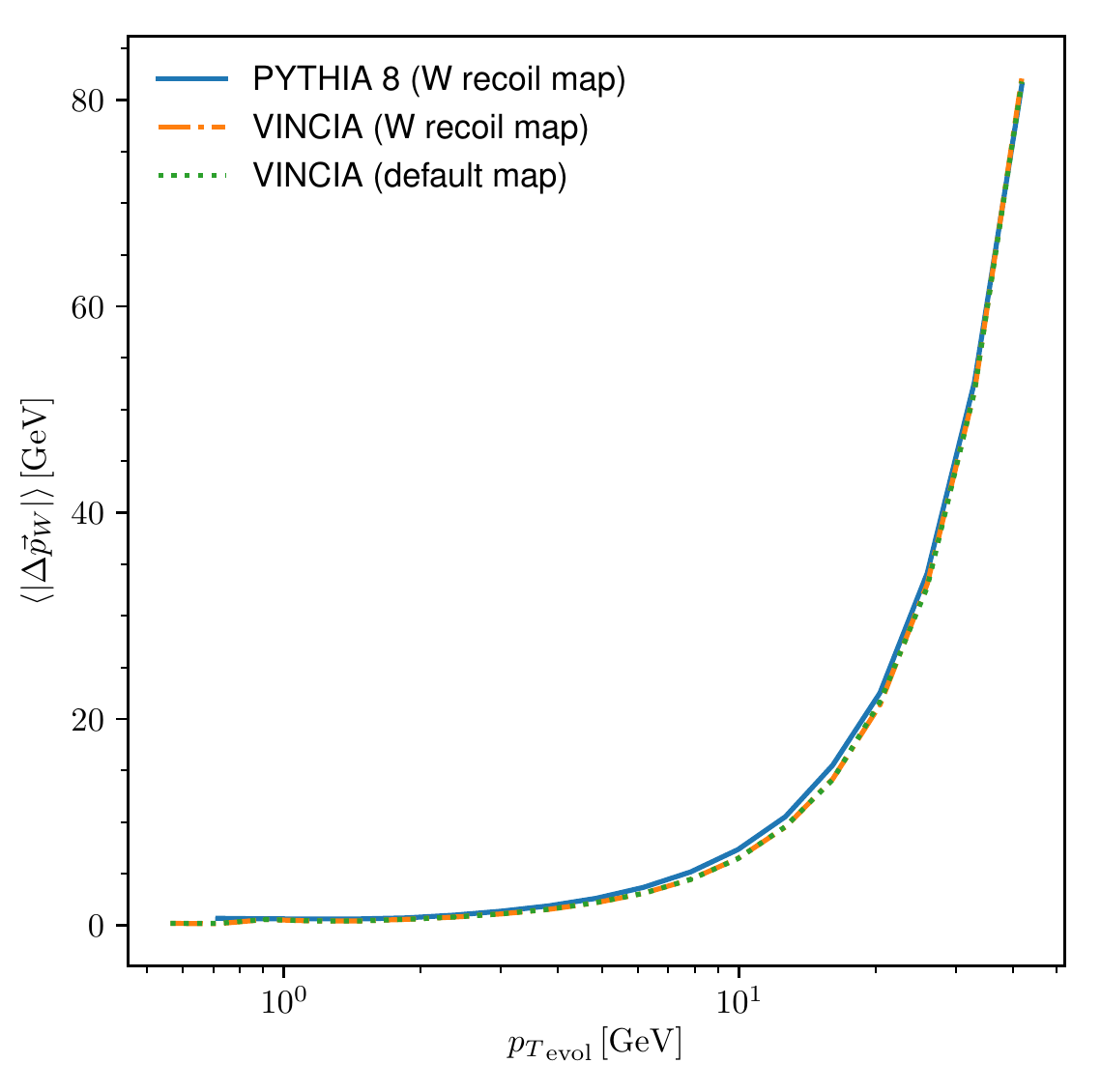}{}
  \caption{}
  \label{fig:recoilinptafterone}
 \end{subfigure}
 \begin{subfigure}{0.4\textwidth}
 \centering
 \includegraphics[width=\textwidth]{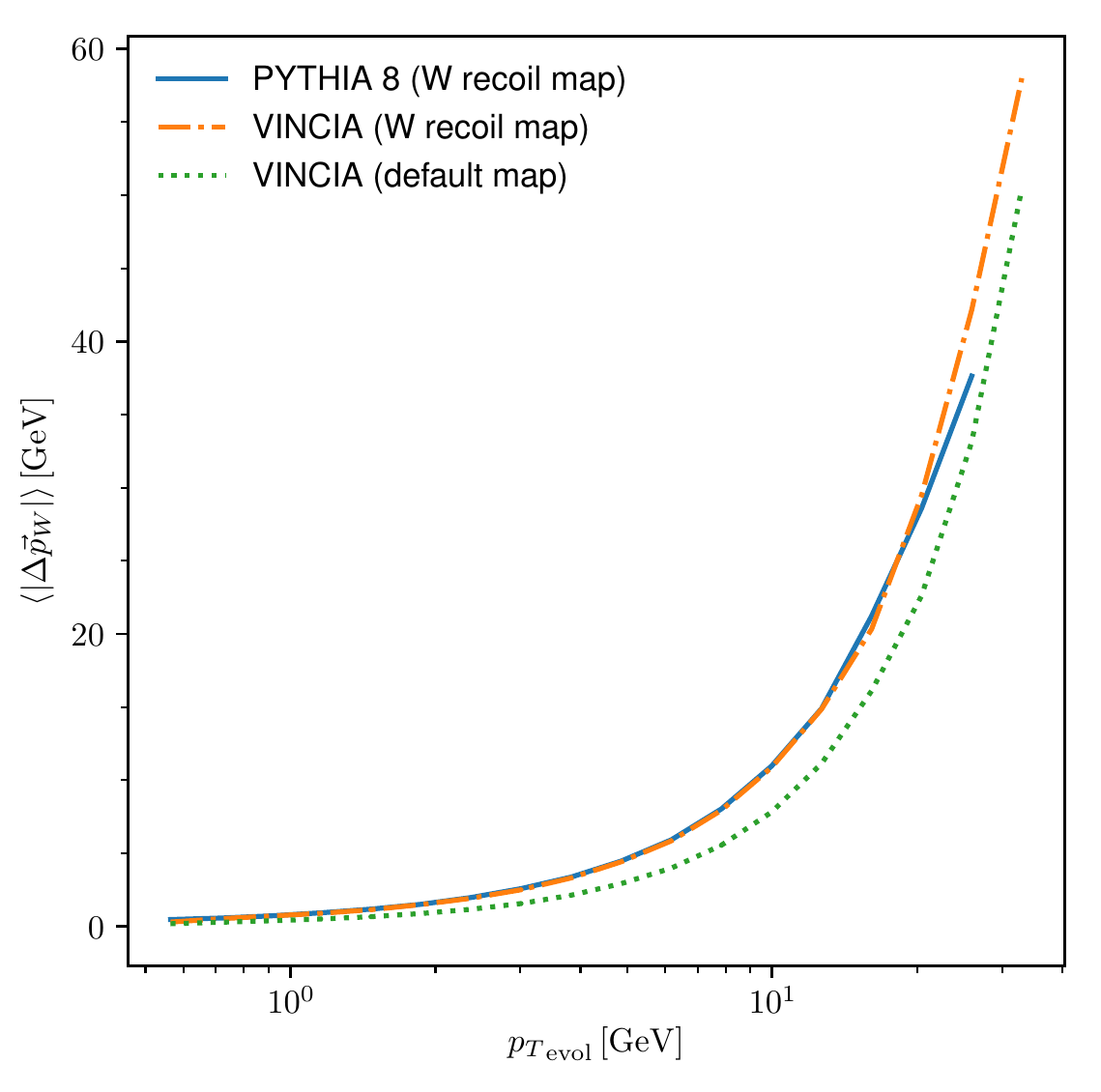}{}
  \caption{}
 \label{fig:recoilinptaftertwo}
 \end{subfigure}
\caption{Plots showing the average change in the three-momentum of the $W$ boson before and after the first emission
\subref{fig:recoilinptafterone}, and between the first and second emission \subref{fig:recoilinptaftertwo}, from the resonance-final antenna in $t\rightarrow bWX$ decays.
This is shown as a function of \py's evolution variable, ${p_T}_\mathrm{evol}$.}
\label{fig:kinmapplots}
\end{figure} 

We shall here validate the kinematic map described in \cref{sec:factorisation}.
We consider the process $e^+e^-\rightarrow t \bar{t} \rightarrow b\bar{b} \ell \bar{\nu}_\ell \bar{\ell} \nu_\ell$
at a hypothetical collider with $\sqrt{s}=1$ TeV. As a measure of the impact of the kinematic map we consider the 
difference in the three-momentum of the $W$ boson, $|\Delta \vec{p}_W|$,  before and after the first and second emissions from resonance decay system.
For \vin~ we compare our default map, where the recoil is shared between all particles in the resonance decay system, and the choice 
where the $W$ boson takes all the recoil; we also compare against \py~ where the latter recoil strategy is activated via the 
setting:

\begin{lstlisting}
 TimeShower:recoilToColoured  = off
\end{lstlisting}

Since we only care about the impact of the kinematic map, we turn off matrix-element corrections in \py~ with:
\begin{lstlisting}
TimeShower:MEcorrections = off
\end{lstlisting}
(\vin~ does not have built-in matrix element corrections at the present time).

In order to have better control over the phase space available for emissions in the decay, for these tests we set the nominal width of both the $t$ quark  and $W$ boson to be zero so that they are all produced at their on-shell masses, i.e.\ with no Breit-Wigner smearing. In addition we turn off the shower prior to decay with 
\begin{lstlisting}
PartonLevel:ISR = off
PartonLevel:FSRinProcess = off
\end{lstlisting}
and veto final-final emissions that occur
prior to the second emission from the resonance-final dipole/antenna using the \texttt{UserHooks} interface. Finally we turn off the QED shower with:
\begin{lstlisting}
PDF:lepton = off
TimeShower:QEDshowerByQ = off
TimeShower:QEDshowerByL = off
TimeShower:QEDshowerByGamma = off
TimeShower:QEDshowerByOther = off
\end{lstlisting}
for \py, and 
\begin{lstlisting}
Vincia:doQED = off
\end{lstlisting}
for \vin.

A plot of the distribution in $|\Delta \vec{p}_W|$ reveals surprisingly large differences between \py~ and \vin.
This effect already appears after the first emission, as can be seen from \cref{fig:recoilafterone};
since there is only the $W$ to absorb the recoil it would be expected that both maps behave the same.
While this is true for the two maps in \vin, a notably harder distribution is given by \py.

After the second emission, shown in \cref{fig:recoilaftertwo}, the discrepancy between the two generators
where the $W$ recoil strategy is employed is less severe, although now \py~ exhibits an earlier drop off. The recoil spectrum for \vin~ 
where the recoil is now shared between the $W$ and the first emission is softer, as would be expected.

In fact, the discrepancies in both plots can be explained when the differences in how phase space is sampled, that 
arise from slight differences in the Sudakov factors, is taken into account.
These sampling differences may be removed by plotting the average  $|\Delta \vec{p}_W|$ as a function of \py's evolution variable,
${p_T}_\mathrm{evol}$ \cite{Sjostrand:2004ef}. (We describe how this is calculated for \vin~ in \cref{app:ptevolCalc}.)
Since this variable is a good representative for the hardness of the emission, it should correlate well
with the amount of recoil required; by averaging, any bias due to different sampling in ${p_T}_\mathrm{evol}$ is removed. 

Indeed, as shown in \cref{fig:recoilinptafterone}  there is virtually no difference between the generators after the first emission. 
After the second emission, shown in \cref{fig:recoilinptaftertwo}, there is agreement for the two generators for the (non-default) case when only the $W$ takes the recoil;
for \vin's default map, there is a softer recoil spectrum as would be expected, since the recoil must now be shared with the first emission. (Whereas for \py's default map, the $W$ receives no recoil from the second emission.)
 
We note that \py~ does not populate the region of high ${p_T}_\mathrm{evol}$ for the second emission. While the maximum value for ${p_T}_\mathrm{evol}$ 
should be determined by the physical phase space, it seems that the region corresponding to large $s_{jk}$ is sampled less efficiently for \py.
This produces the drop off seen in \cref{fig:recoilaftertwo}.

\begin{figure}[t]
\centering
 \includegraphics[width=0.4\textwidth]{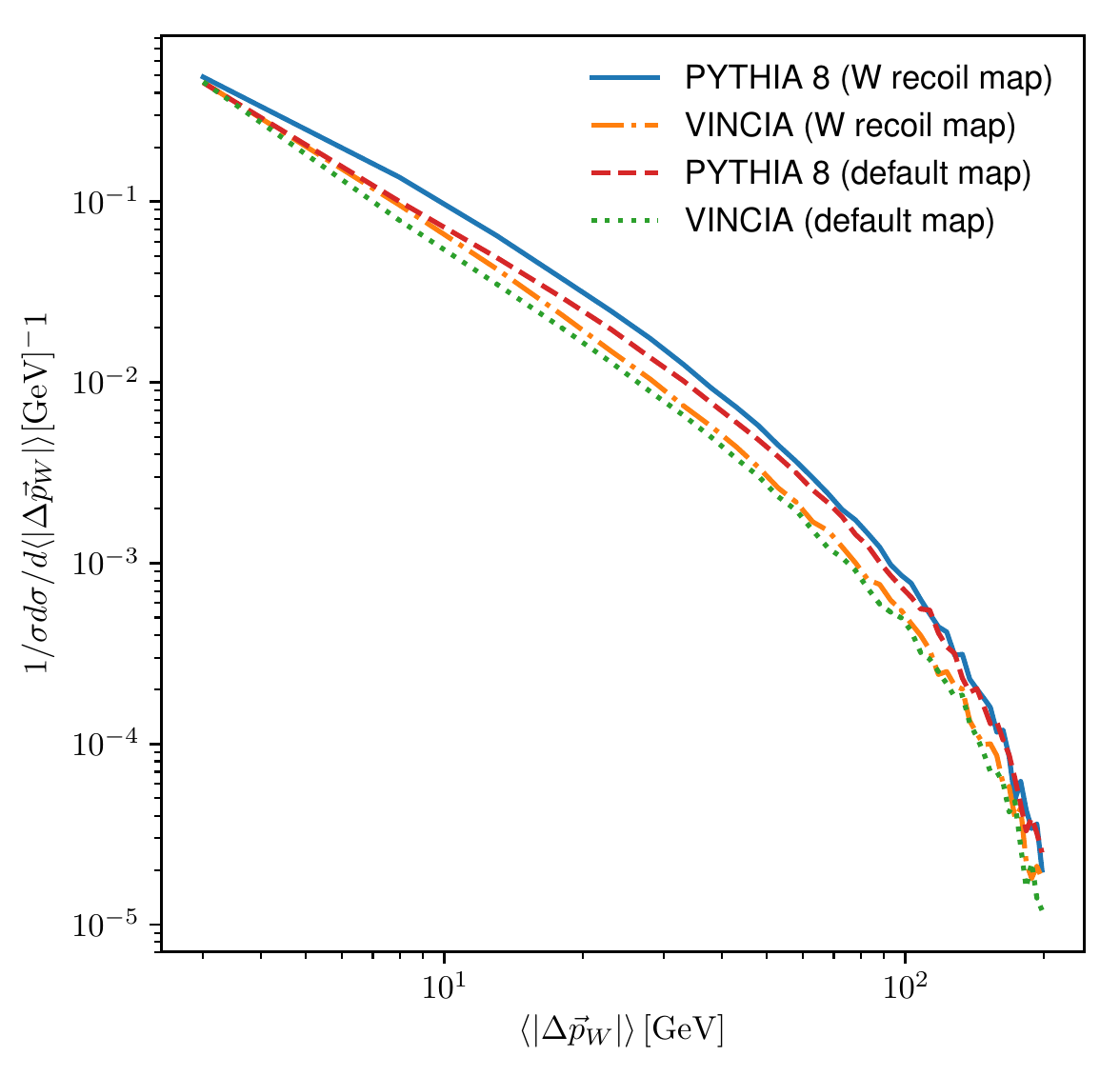}{}
  \caption{Plot showing the distribution in the change in three-momentum between the $W$ boson at Born-level and post-shower.}
 \label{fig:recoilfinal} 
\end{figure}

Finally we show the distribution in the three-momentum between the $W$ boson at Born-level and post-shower in \cref{fig:recoilfinal}, 
including the default option for \py. Generally, \py~ gives more recoil than \vin~, and for the largest recoils, both \py~ maps converge.
This is consistent with \py~ generating a harder first emission than \vin. 
(This conclusion changes when matrix-element corrections are turned on: now \vin~ gives a harder first emission.)
For both \py~ and \vin~ the $W$ recoil map produces slightly more recoil than the map without, although the effect is more subtle in \vin.

\subsection{Coherence}
\label{sec:coherence}

\subsubsection{Coherence in Production}

It was noted in \cite{Skands:2012mm} that for $p\bar{p}\rightarrow t\bar{t}$ at the Tevatron, parton showers that exhibit coherence
in  initial-final dipoles, are capable of producing  non-zero forward-backwards asymmetries, defined  (differentially in a generic observable $\mathcal{O}$) as:
\begin{equation}
 A_{FB}(\mathcal{O}) = \frac{\left.\frac{\mathrm{d}\sigma}{\mathrm{d}\mathcal{O}}\right|_{\Delta y>0} - \left.\frac{\mathrm{d}\sigma}{\mathrm{d}\mathcal{O}}\right|_{\Delta y<0}}
 {\left.\frac{\mathrm{d}\sigma}{\mathrm{d}\mathcal{O}}\right|_{\Delta y>0}+\left.\frac{\mathrm{d}\sigma}{\mathrm{d}\mathcal{O}}\right|_{\Delta y<0}}
\end{equation}

Conceptually, this occurs because the initial-final colour antennae span a much greater angle when the outgoing top is going backwards relative to the direction of the corresponding 
incoming coloured parton (which tends to be a valence quark at the Tevatron and hence correlated with the beam direction) than when it is going forwards.
Initial-final antennae with forward-going tops hence do not radiate as much as ones with backwards-going tops, which shows up as a net positive asymmetry
(more forwards-going tops than backwards-going ones) at low values of the transverse momentum of the $t\bar{t}$ system, $p_{T}(t\bar{t})$,
and a net negative one at high values. 

We reproduce the analysis that was implemented in \cite{Skands:2012mm} using \rivet~\cite{Buckley:2010ar}, for $p\bar{p}\rightarrow t\bar{t}$ with $\sqrt{s}=1.96$ TeV.
Here we are concerned with coherence in production, thus for the purpose of this analysis we prevent the top from decaying and perform the analysis
at Monte Carlo ``truth''-level (that is, we assume that we can perfectly identify all final state partons).
Nevertheless this still constitutes a good validation of the antennae given in \cref{sec:antennae} as the same set are used for both the initial-final and resonance final cases
(albeit with explicitly massless initial partons in the former).

\begin{figure}[t]
\centering
\includegraphics[width=0.45\textwidth]{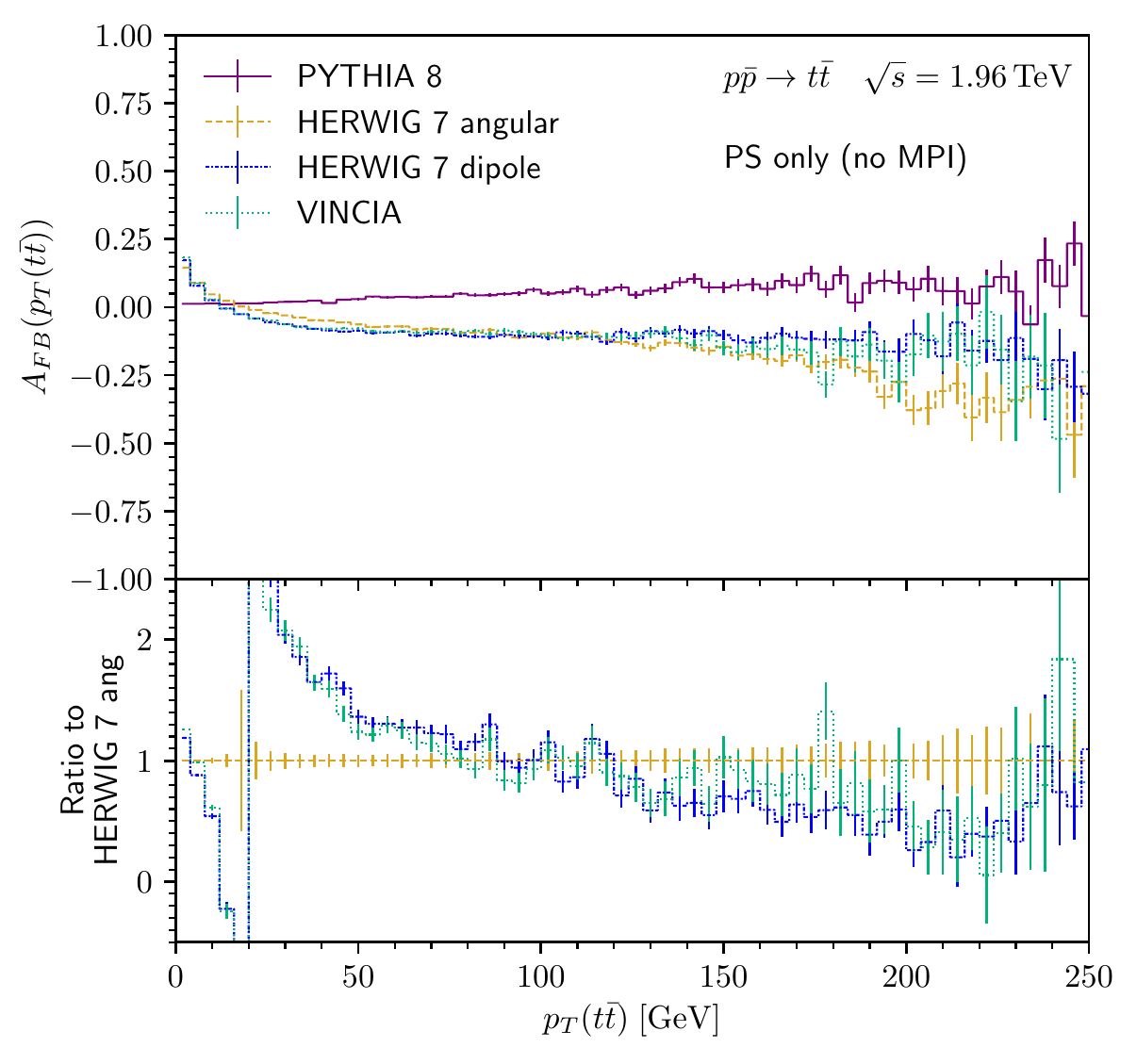}
\caption{Plot showing the forwards-backwards asymmetry as a function of the transverse momentum of the $t \bar{t}$ system
for the 1.96 TeV Tevatron $p\bar{p}$ collider.}
\label{fig:ppasymm}
\end{figure}

In \cref{fig:ppasymm} we show the differential distribution of the forwards-backwards asymmetry as a function of $p_T(t\bar{t})$. Notably \py~8 does not predict an asymmetry, remaining close to zero and essentially flat across the range of 
transverse momentum. On the other hand the coherent showers, namely both the \herwig~7 showers (angular-ordered and Catani-Seymour dipole)  and \vin~ qualitatively exhibit similar
dependence of the asymmetry on $p_T(t\bar{t})$: the asymmetry is small and positive for small values of $p_T(t\bar{t})$, and becomes negative for larger $p_T(t\bar{t})$. In the bottom panel
we show the ratio of the three coherent showers to the \herwig~7 angular-ordered shower. The distribution for \vin~ has a very similar shape to 
the \herwig~7 dipole shower, starting with slightly higher positive asymmetry for small $p_T(t\bar{t})$, before dropping more steeply to negative values and flattening off, relative to the angular-ordered shower.

Finally we note that it is also possible to produce an integrated asymmetry.
One might expect that unitarity should imply the asymmetry integrates to zero, since the shower starts from the LO Born cross section (which does not have an asymmetry) and the inclusive cross section is preserved.
However because recoils in the shower can change the relative ordering of the $t\bar{t}$ pair, there can be sufficient migration between bins of rapidity to 
produce an integrated asymmetry. For \vin~ this is at the level of $~7\%$, in line with the other coherent showers  (and compared to 
$~6\%$ predicted by the first non-trivial leading-order QCD prediction that yields an asymmetry).

\begin{figure}[t]
\centering
 \begin{subfigure}{0.45\textwidth}
 \centering
 \includegraphics[width=\textwidth]{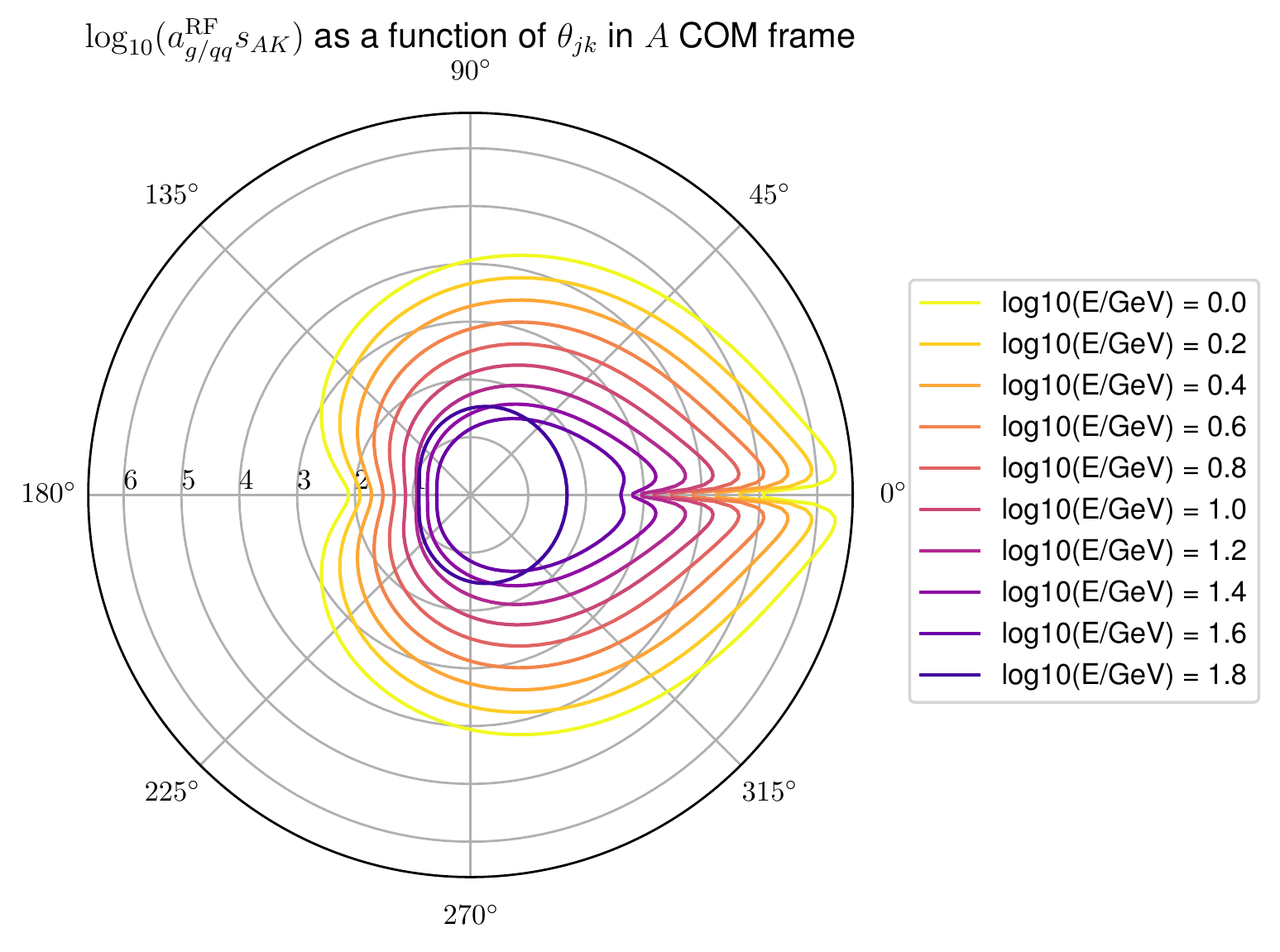}{}
  \caption{}
  \label{fig:rfant-com}
 \end{subfigure}
 \begin{subfigure}{0.45\textwidth}
 \centering
 \includegraphics[width=\textwidth]{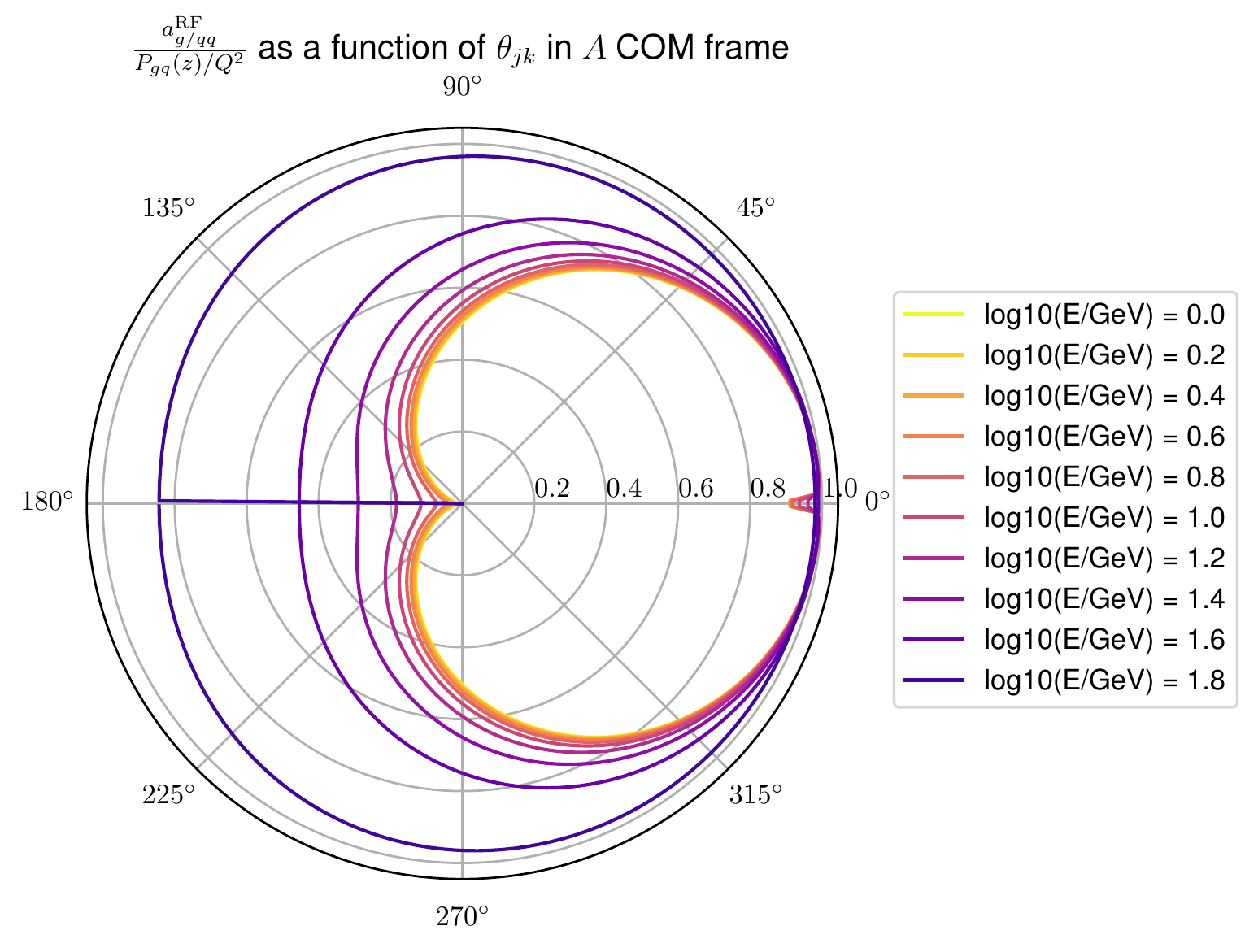}{}
  \caption{}
  \label{fig:rfant-ratio}
 \end{subfigure}
\caption{Plots showing the antenna function $a^{RF}_{g/qq}$ for  $t\rightarrow b W$ in the top quark centre-of-mass frame. The polar angle corresponds to
the opening angle between the $b$ quark (after branching) and the gluon emission; in \cref{fig:rfant-com} the radial coordinate corresponds to $\log_{10}(a^{RF}_{g/qq}s_{AK})$
while in \cref{fig:rfant-ratio} it corresponds to the ratio to the Altarelli-Parisi splitting function. The different contours correspond to different gluon energies, equally
spaced on a logarithmic scale. }
\label{fig:rfant-coherence}
\end{figure} 

\subsubsection{Coherence in Decay}

We proceed now to consider coherence in decay of the resonance. In \cref{fig:rfant-coherence}
we evaluate the antenna function in \cref{eq:antrf_qq} for $t\rightarrow b W$, and use its value as the radial 
coordinate in a polar plot. The polar angle corresponds to the opening angle between the $b$ quark (after branching) and the gluon emission
in the centre-of mass frame of the top, with the original $b$ quark oriented at 0 degrees.
We do this for fixed values of the energy of the gluon (again, in the top centre of mass frame), evenly spaced on a logarithmic scale.

We indicate
the value of the energy using the colour, with paler yellow corresponding to soft emissions, and darker purple corresponding to harder emissions. 
In \cref{fig:rfant-com} we plot the logarithm of the antenna function, multiplied by $s_{AK}$ to give a dimensionless value.
We can clearly see that both soft and quasi-collinear emissions are logarithmically enhanced, with the suppression in the forwards direction corresponding to the
well-known mass `dead-cone' effect \cite{Ellis:1991qj}. 

In \cref{fig:rfant-ratio} we instead show the ratio of the antenna function to the Altarelli-Parisi splitting function $P_{g\rightarrow gq}(z)/Q^2$ (taking $Q^2 = s_{jk}$ and $z = s_{ak}/s_{AK}$).
For quasi-collinear emissions (i.e. in the forwards direction) for all emission energies, the ratio tends to one (the slight deviation from this inside the mass-cone corresponds to the two results
tending mutually to zero at slightly different rates due to the presence of additional finite terms for the antenna function).
However in the away region we see that the (coherent) antenna pattern is strongly suppressed relative to that of the (incoherent) Dokshitzer-Gribov-Lipatov-Altarelli-Parisi kernel. This is therefore quite a good visualisation of the impact of coherence in decay.

\subsection{B-jet Profiles}
\label{sec:bjetprofiles}

We can systematically investigate the combined effect of the kinematic map and of coherence by examining the impact upon 
the shape of $b$-jets in $t\bar{t}$ production at the LHC.  
Specifically we consider the jet-profile \cite{Aad:2011kq}, defined as 
\begin{equation}
   \rho(r)=\frac 1 {\Delta r} \frac 1
  {N_{\mathrm{jets}}}\sum_{\mathrm{jets}}\frac{p_\perp(r-\Delta r/2,r+\Delta r/2)}{p_\perp(0,R)},
 \label{eq:profile}
\end{equation}
This variable represents the proportion of the jet's transverse momentum that is carried by the particles inside an annulus of radius $r = \sqrt{(\Delta y)^2 + (\Delta \phi)^2}$, with (bin) width $\Delta r$. 
It is thus a measure of how the momentum is distributed throughout the jet.
Where more wide-angle radiation is produced,
this should give rise to a broader jet profile.

Our treatment, implemented using \rivet, is similar to the analysis in \cite{Aad:2011kq}, however we only consider the two hardest $b$-tagged jets. 
(This corresponds to the jet containing a $b$ quark at parton level, and a $b$ hadron for particle-level analysis, that is, we do not trace the 
history of identified particles through the event record).
We take $\sqrt{s}=13$ TeV; jets are constructed using the anti-$k_T$ algorithm \cite{Cacciari:2008gp} with $R=0.6$, as implemented in \fastjet~\cite{Cacciari:2011ma}.

\begin{figure}[t]
\centering
 \begin{subfigure}{0.45\textwidth}
 \centering
 \includegraphics[width=\textwidth]{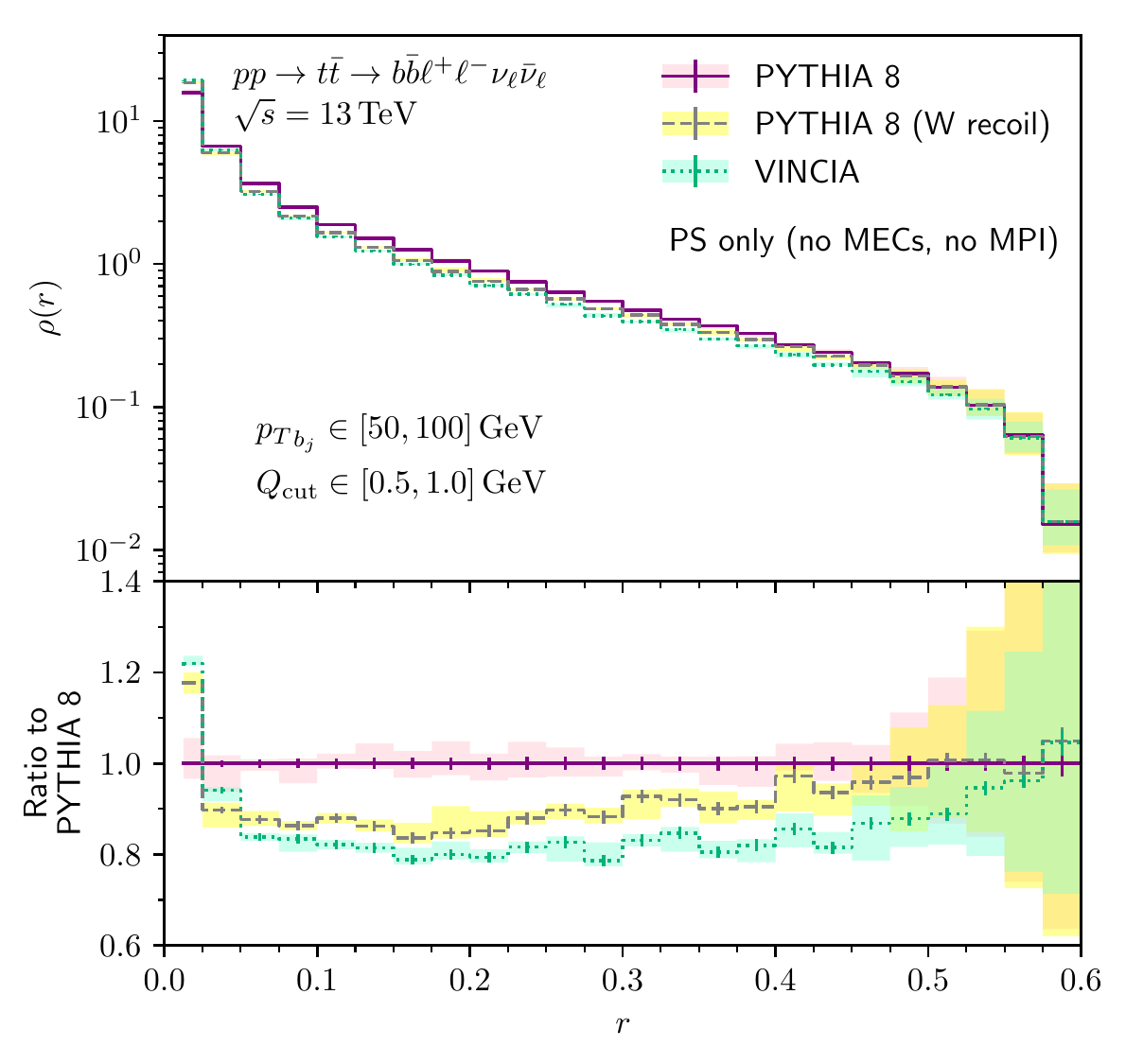}{}
  \caption{}
    \label{fig:bprofps_nomecs}
 \end{subfigure}
 \begin{subfigure}{0.45\textwidth}
 \centering
 \includegraphics[width=\textwidth]{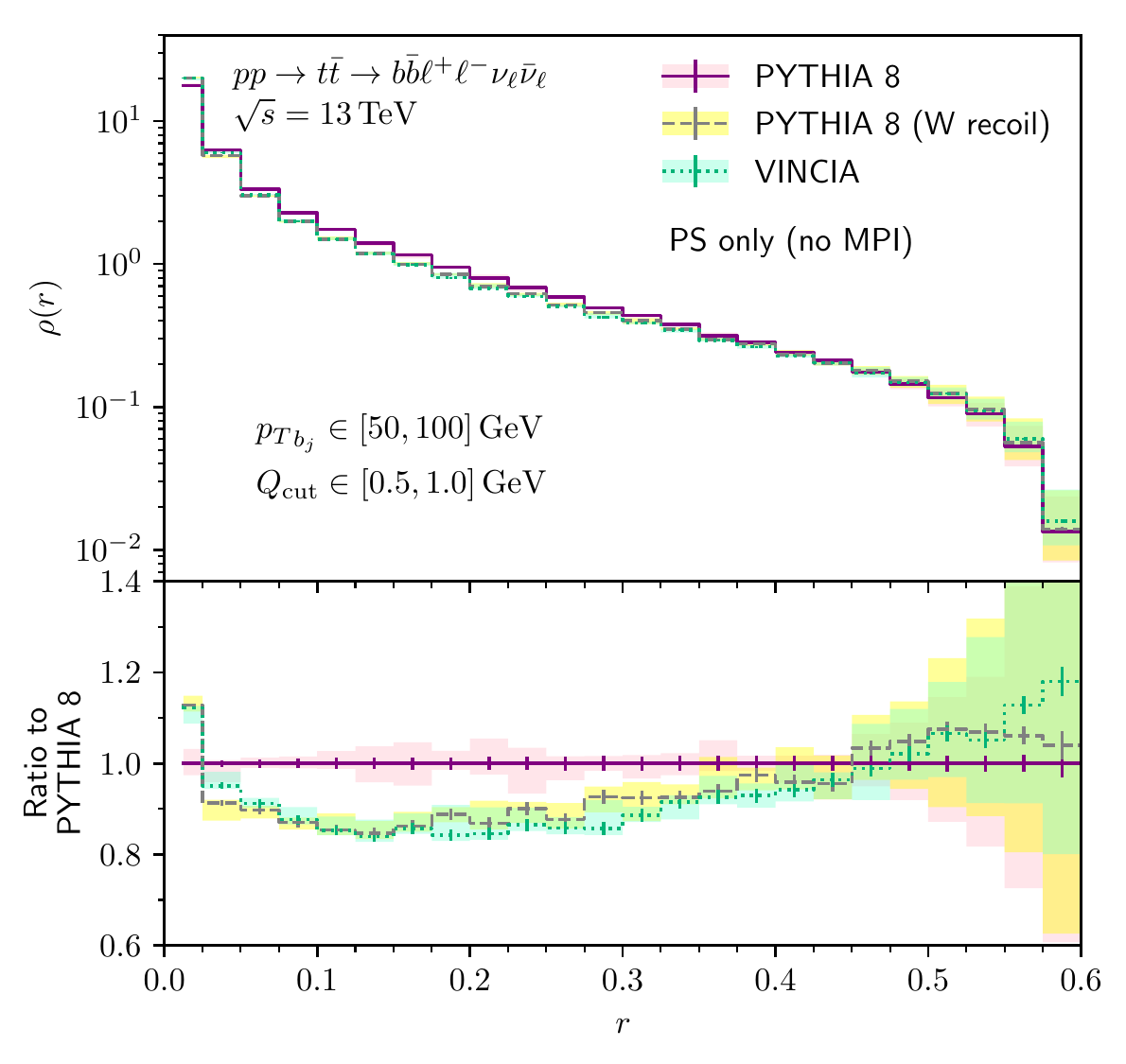}{}
  \caption{}
  \label{fig:bprofps}
 \end{subfigure}
 \caption{Distribution of the jet profile (as defined in \cref{eq:profile}) given by \vin~ and \py~ for a slice of transverse momentum:
$p_T \in [50,100]$ GeV at parton-level (prior to hadronisation and underlying event). Results are shown without \subref{fig:bprofps_nomecs} and with
\subref{fig:bprofps} MECs.}
\label{fig:bjetprofiles}
\vspace{-1em}
 \end{figure} 

 \begin{figure}[t] 
 \centering
 \begin{subfigure}{0.45\textwidth}
 \centering
 \includegraphics[width=\textwidth]{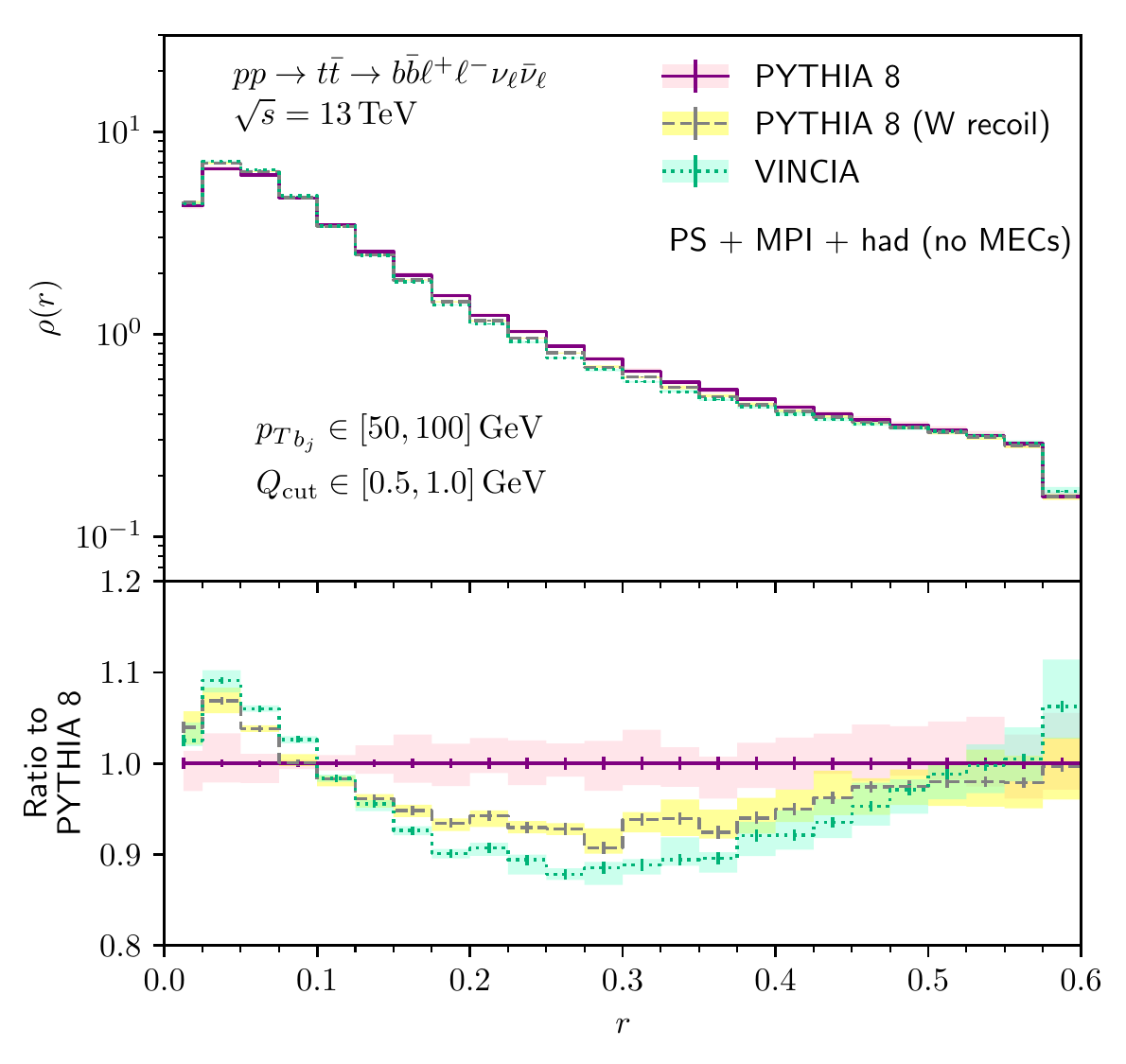}{}
  \caption{}
\label{fig:bprofmpihad_nomecs}
 \end{subfigure}
 \begin{subfigure}{0.45\textwidth}
 \centering
 \includegraphics[width=\textwidth]{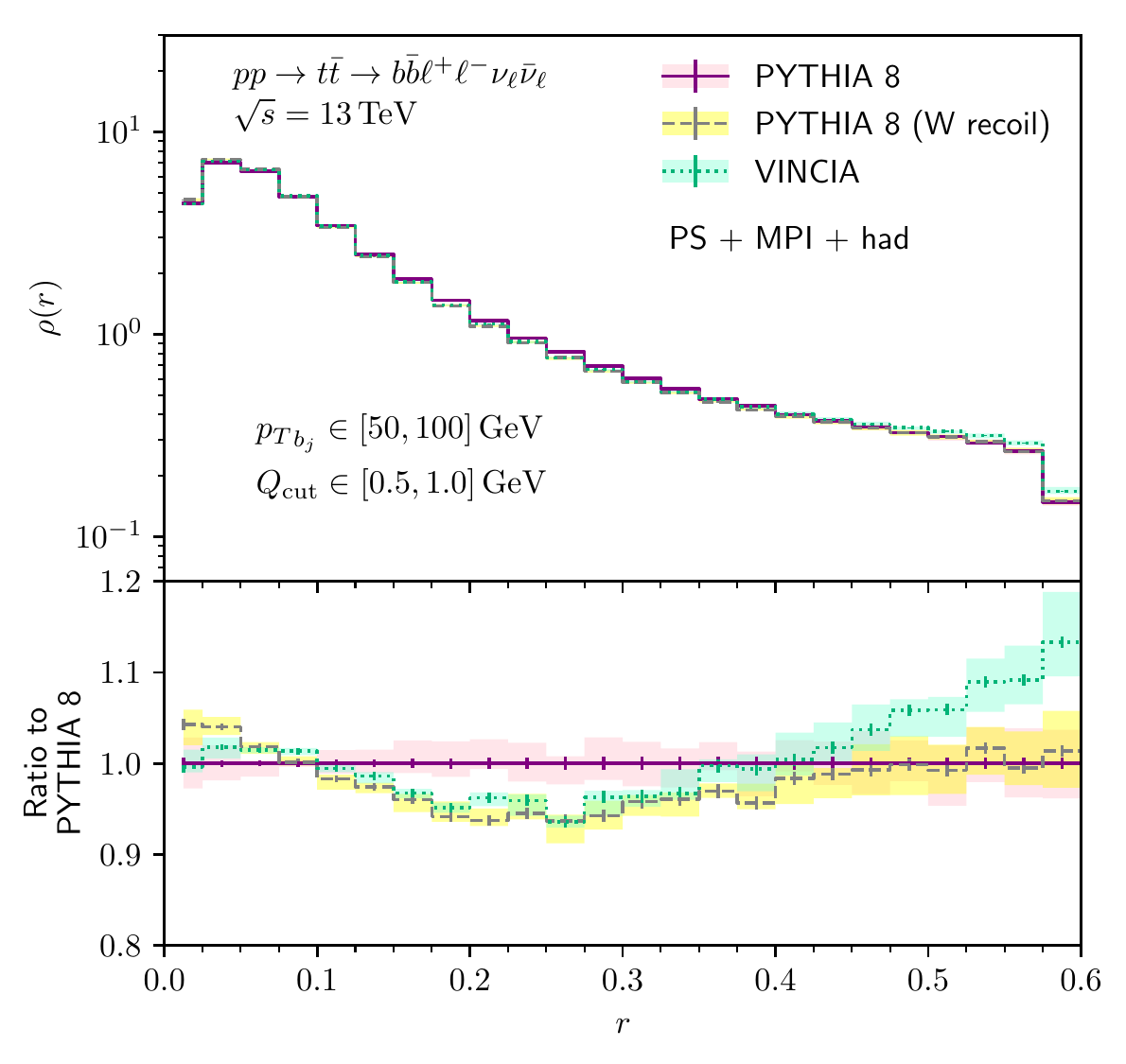}{}
  \caption{}
  \label{fig:bprofmpihad}
 \end{subfigure}
\caption{As for  \cref{fig:bjetprofiles} but now with hadronisation and underlying event included.
Results are shown without \subref{fig:bprofmpihad_nomecs} and with
\subref{fig:bprofmpihad} MECs.}
\label{fig:bjetprofiles_had}
\end{figure} 

\begin{figure}[t]
 \centering
 \includegraphics[width=0.45\textwidth]{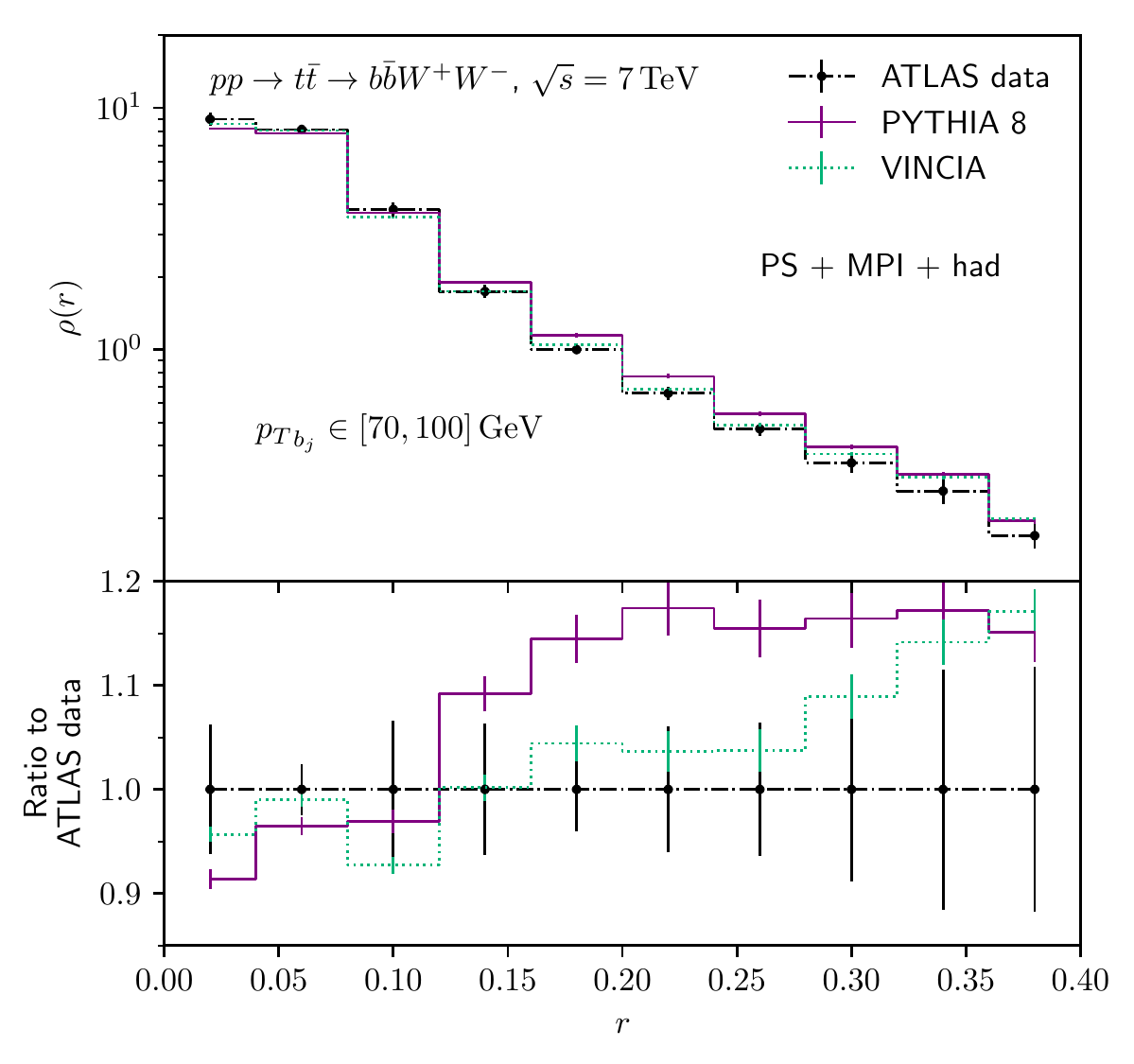}
 \caption{Distribution of the $b$-jet profile as measured by ATLAS in \cite{Aad:2013fba} for $t\bar{t}$ production at $\sqrt{s} = 7$ TeV.
 }
 \label{fig:bprofile_atlas}
\end{figure}

In \cref{fig:bjetprofiles}  we show the distribution in $\rho(r)$ as given by \vin~ and \py~ for a slice of jet transverse momentum $p_T \in [50,100]$ GeV,
where the simulation of QED, underlying event and hadronisation has been turned off. The shaded bands corresponds to varying the cutoff scale for each
shower in the interval $Q_\mathrm{cut} \in [0.5, 1]$ GeV 
\footnote{This corresponds to varying \texttt{Vincia:cutoffScaleFF}, \texttt{Vincia:cutoffScaleII}, and \texttt{Vincia:cutoffScaleIF}
for \vin, and \texttt{TimeShower:pTmin} and \texttt{SpaceShower:pTmin} in \py~. 
In principle we should also vary the regularisation scale \texttt{SpaceShower:pT0Ref}, 
but since the effects described here are dominated by the final state shower, we expect the impact to be fairly minimal.
}
(and the central line corresponds to $Q_\mathrm{cut} = 0.75$ GeV).

By default \py~ includes MECs; in \cref{fig:bprofps_nomecs} these have been turned off, such that the splitting kernels are just the basic Altarelli-Parisi ones.
Here \vin~ has a significantly more narrow $b$-jet profile than \py.
We find that this can be in part, but not fully, accounted for by \py's alternative recoil strategy where the $W$ always takes the recoil. 
(Both choices of recoil strategy in \vin~ perform similarly and therefore here we only show the default option.) 
When a coloured parton inside the $b$-jet receives the recoil, this has more potential to ``kick'' particles around inside the jet cone 
than a strategy where the $W$ boson receives all or most of recoil, thereby broadening the spectrum.
We interpret the remaining difference as due to coherence. In the centre-of-mass frame, \vin's coherent antenna pattern suppresses emissions in the backwards direction; after boosting this should be manifest as a suppression of 
wide-angle radiation, giving rise to narrower jets. 

In \cref{fig:bprofps}, we include MECs for \py; now the alternative kinematic map fully accounts for the difference between \vin and default \py. 
From this we conclude that \vin's antenna functions perform very similarly to \py~ with MECs. To put this another way, including MECs in \py~ effectively recovers the missing coherence in the radiation pattern while the difference caused by the different recoil strategies persists. 

In \cref{fig:bjetprofiles_had} we show the same distribution, now with hadronisation and underlying event included. Although there is 
a significant broadening of the profiles, including some suppression in the central region, the same qualitative differences remain 
(albeit reduced in size - note the change in scale). It should be noted that although MPI and hadronisation are in all cases handled by \py,  
the default values used for those models by \vin~\cite{Fischer:2016vfv} are in general different from those in the Monash tune \cite{Skands:2014pea} used by \py~8.2. 
However, we do not find that our results change qualitatively even if \vin\ is forced to use the Monash parameters.

The choice of kinematic map can be regarded as a theoretical uncertainty, which  may have an impact on the reconstructed top mass (studied in more detail below)
as well as on the efficiency of $b$-taggers used by ATLAS and CMS.
However, given the stability of our observations to non-perturbative effects we suppose that it might be physically measurable;
in-situ measurements of $b$-jet profiles such as those in \cite{Aad:2013fba}
may provide insight into which kinematic map is most physical and provide constraints for tuning and uncertainty evaluations.
We do not re-tune here, but merely note that when we repeat the analysis of \cite{Aad:2013fba} using \rivet,
we do observe the same qualitative results. 
This is demonstrated in \cref{fig:bprofile_atlas}, in which we show the $b$-jet profile measured by ATLAS for jets of 
transverse momenta $p_T \in [70,100]$ GeV, for the default tunes of \py~ and \vin~.
Despite the experimental uncertainties being fairly large, \vin's narrower jet profile appears to agree better with the data than \py's broader one. (However a stronger conclusion could be drawn if more data were to be collected.)

Finally we comment that both existing $b$-jet substructure measurements \cite{Sirunyan:2018asm}, and measurements of the distribution of the angular separation between the reconstructed top quark 
and additional jets \cite{Sirunyan:2018wem} made by CMS may also be sensitive to the choice of kinematic map.
Thus a systematic study of such observables may help constrain which choice is more physical; however, we defer such a phenomenological study to further work.

\subsection{Parton Shower + Fixed Order comparisons}
\label{sec:psnlo}

Later in this section we will  assess the impact of shower ambiguities on distributions relevant to the 
reconstruction of the top quark mass. We compare between \vin, \py~8.240 and \herwig~7.1.4 (using the angular-ordered shower),
where all parton showers have been matched to NLO accuracy with \pwgii~ according to the method described in \cref{sec:matching}.
Specifically the same input events were used for all parton showers, and hence the inclusive cross section should be identical
in all cases. Furthermore, for all results that follow (both here and in \cref{fig:realistic}) the masses of the top and bottom quark were fixed
across all generators to $m_t = 171$ GeV and $m_b = 4.8$ GeV respectively.

\begin{figure}[t]
 \centering
 \begin{subfigure}{0.45\textwidth}
 \centering
 \includegraphics[width=\textwidth]{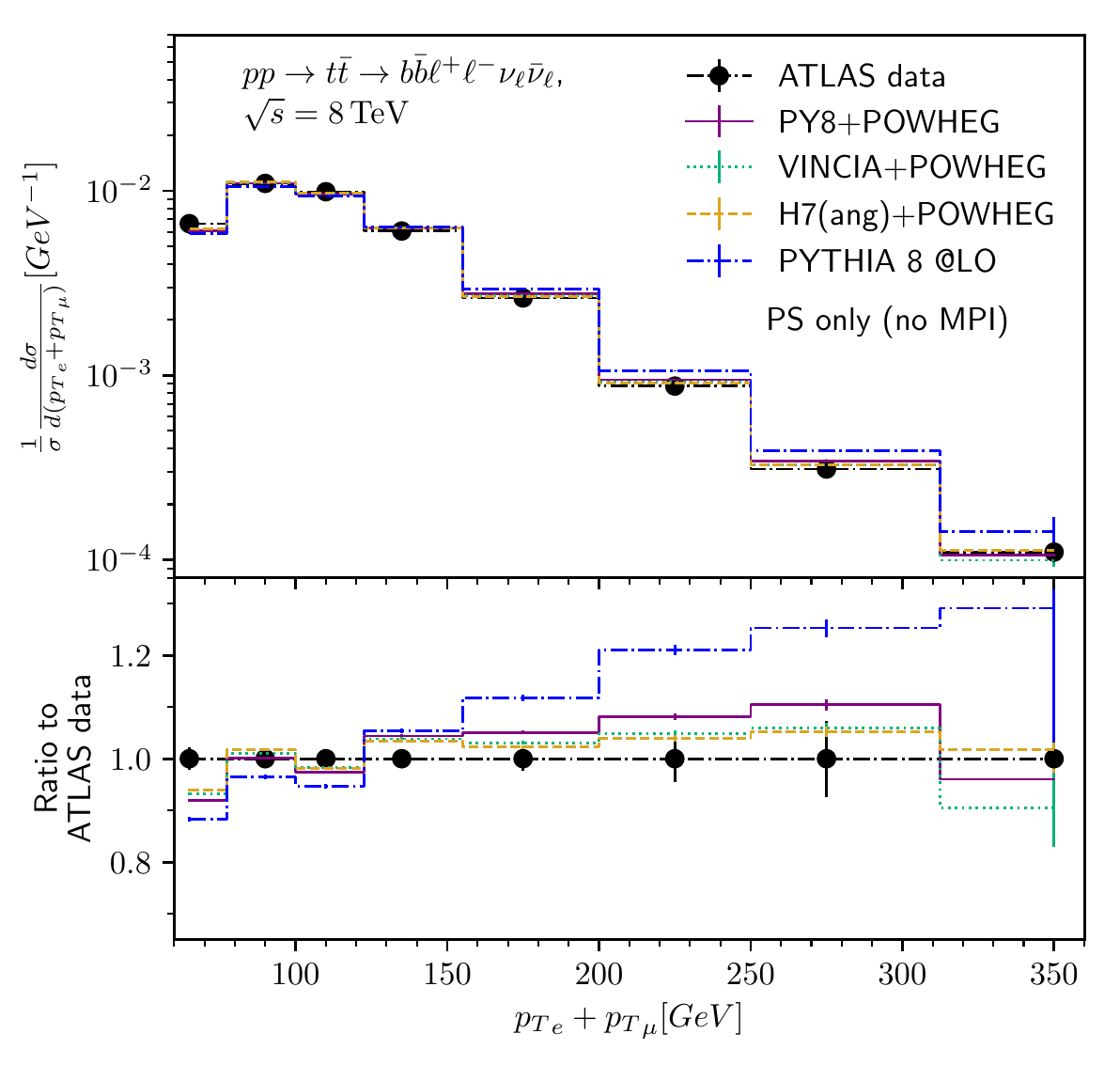}{}
  \caption{}
  \label{fig:pTsumlepton}
 \end{subfigure}
 \begin{subfigure}{0.45\textwidth}
 \centering
 \includegraphics[width=\textwidth]{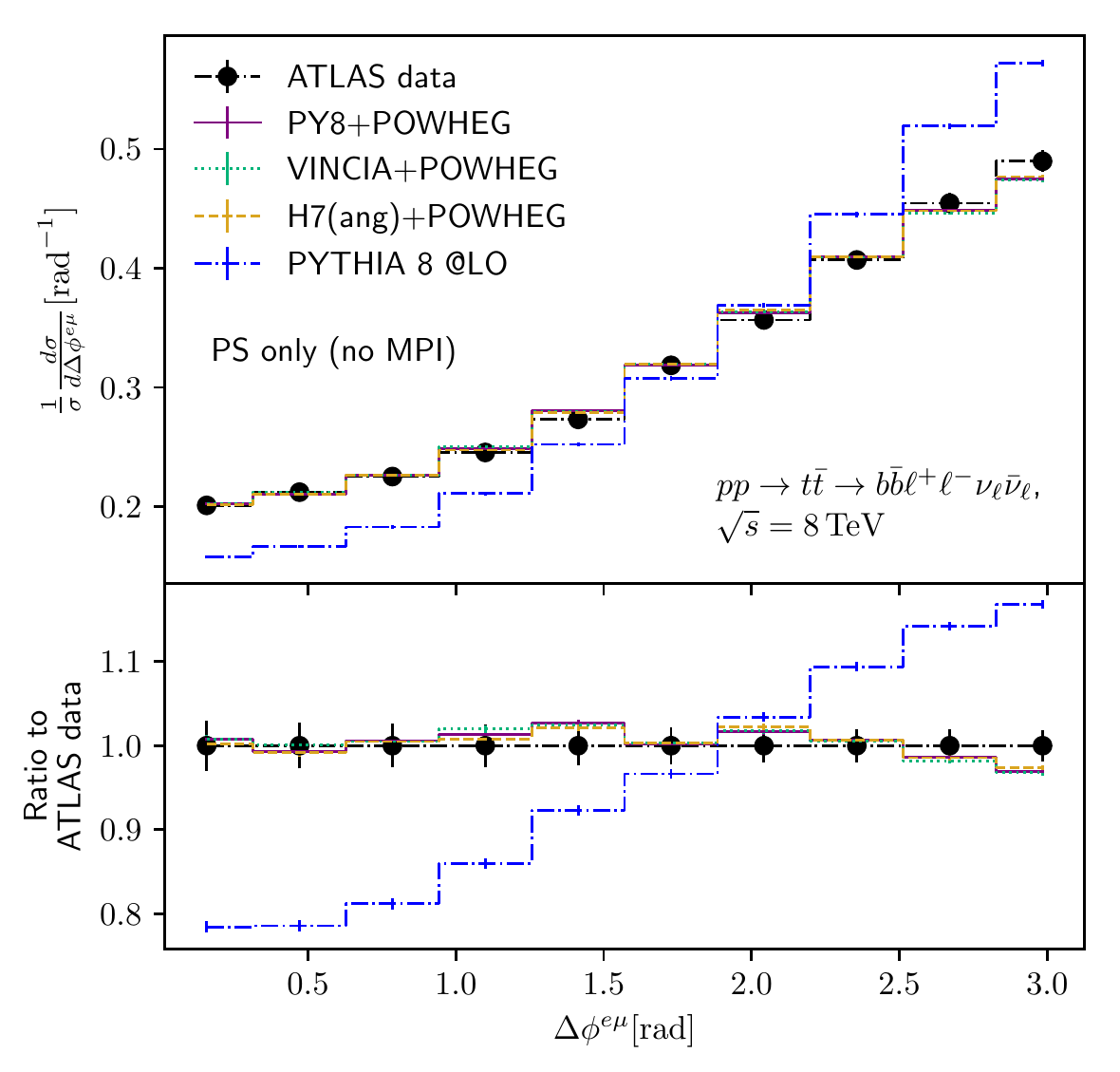}{}
  \caption{}
    \label{fig:dphiempu}
 \end{subfigure}
\caption{Differential cross section with respect to the transverse momentum of the lepton pair, \subref{fig:pTsumlepton}, and 
the azimuthal angle between the lepton pair, \subref{fig:dphiempu}, as measured by ATLAS \cite{Aaboud:2017ujq} in dileptonic $t\bar t$ production at $\sqrt{s}=8$ TeV.
Comparisons are shown between generators matched to NLO accuracy using \pwgii, and LO accuracy using \py~ standalone.}
\label{fig:NLOtestplots}
\end{figure}

As a validation of the matching to NLO, we reproduced an ATLAS analysis which measured differential lepton distributions in 
dileptonic $t\bar t$ production at $\sqrt{s}=8$ TeV \cite{Aaboud:2017ujq}, that was implemented in \rivet.
The distributions in this study were found to be relatively insensitive to the choice of parton shower, but NLO accuracy 
is required for a reasonable description of data. 
In \cref{fig:NLOtestplots} we compare the NLO matched results to data, as well as a leading order plus parton shower 
prediction from \py~8 for the distributions in the sum of transverse momenta of the two leptons, and the 
difference in azimuthal angle between the leptons.
The distributions are normalised to the cross section to highlight the difference in shape in going from leading order to 
next-to-leading order. The leading order predictions show large shape differences with respect to data, while the 
NLO matched predictions are consistent with both data and each other. The level of variation is consistent with
that seen in the original analysis~\cite{Aaboud:2017ujq}.

\begin{figure}[t]
 \centering
 \includegraphics[width=0.45\textwidth]{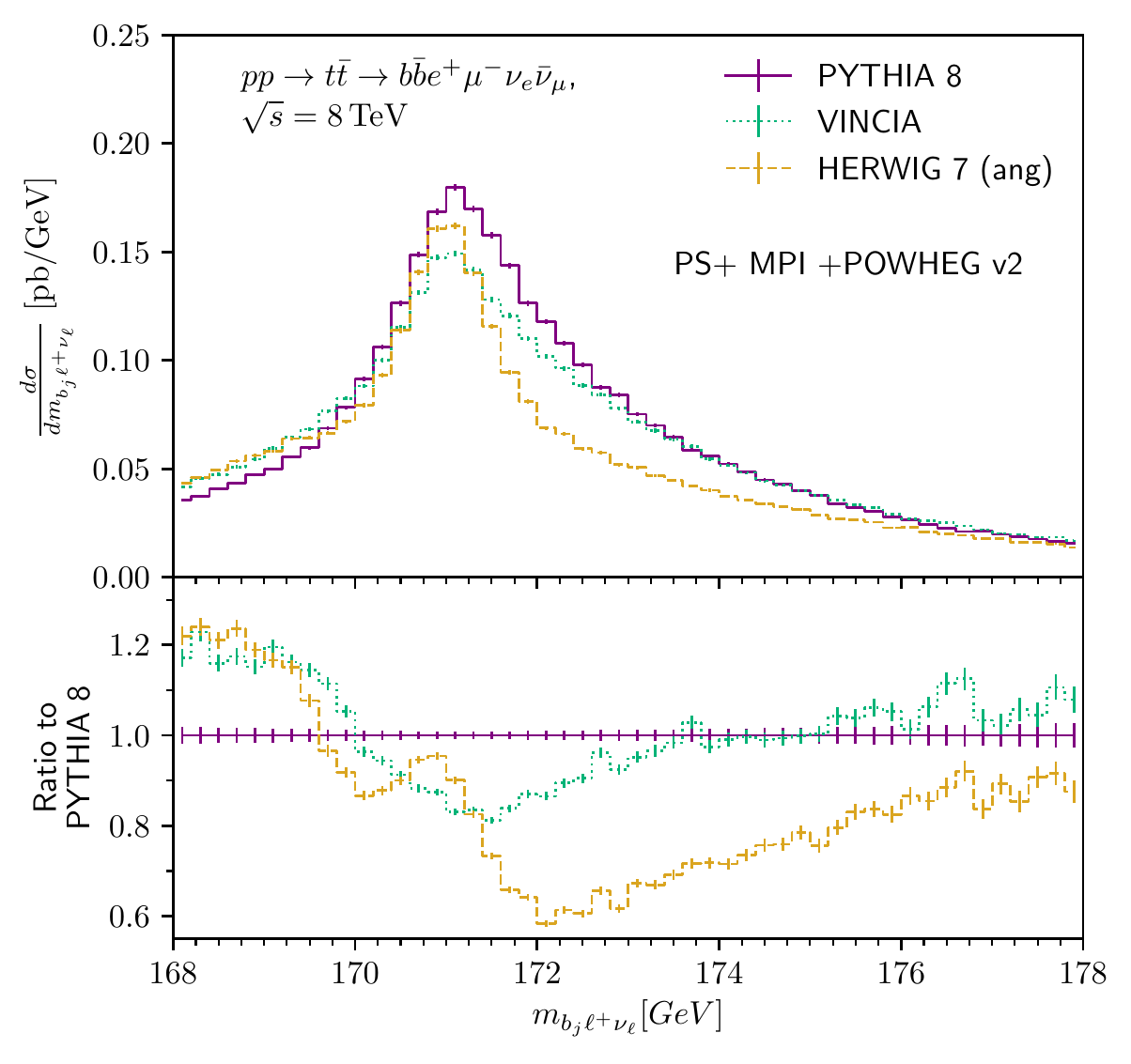}{}
\caption{Plot showing the differential cross section as function of the invariant mass of the $b_j \ell^+ \nu_\ell$ in dileptonic top pair production at the 
LHC with $\sqrt{s}=8$ TeV. Parton shower predictions matched to NLO accuracy using \pwgii~ are compared. Results are shown prior to hadronisation,
but including underlying event (MPI).
}
\label{fig:mblpnu_mpi}
\end{figure}

\begin{figure}[t]
 \centering
 \begin{subfigure}{0.45\textwidth}
 \centering
 \includegraphics[width=\textwidth]{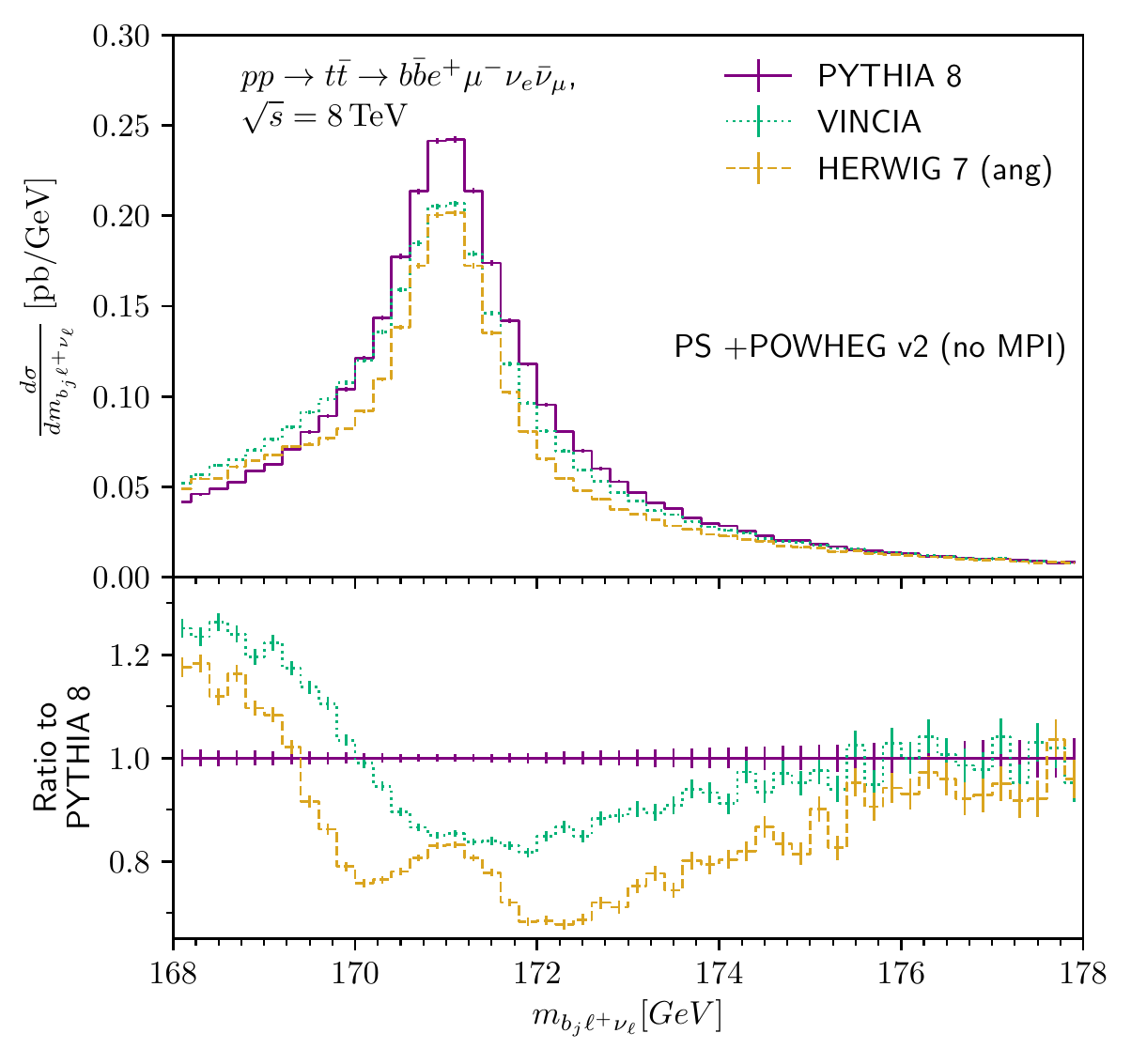}{}
  \caption{}
  \label{fig:mbmlpnu_nompi}
 \end{subfigure}
 \begin{subfigure}{0.45\textwidth}
 \centering
 \includegraphics[width=\textwidth]{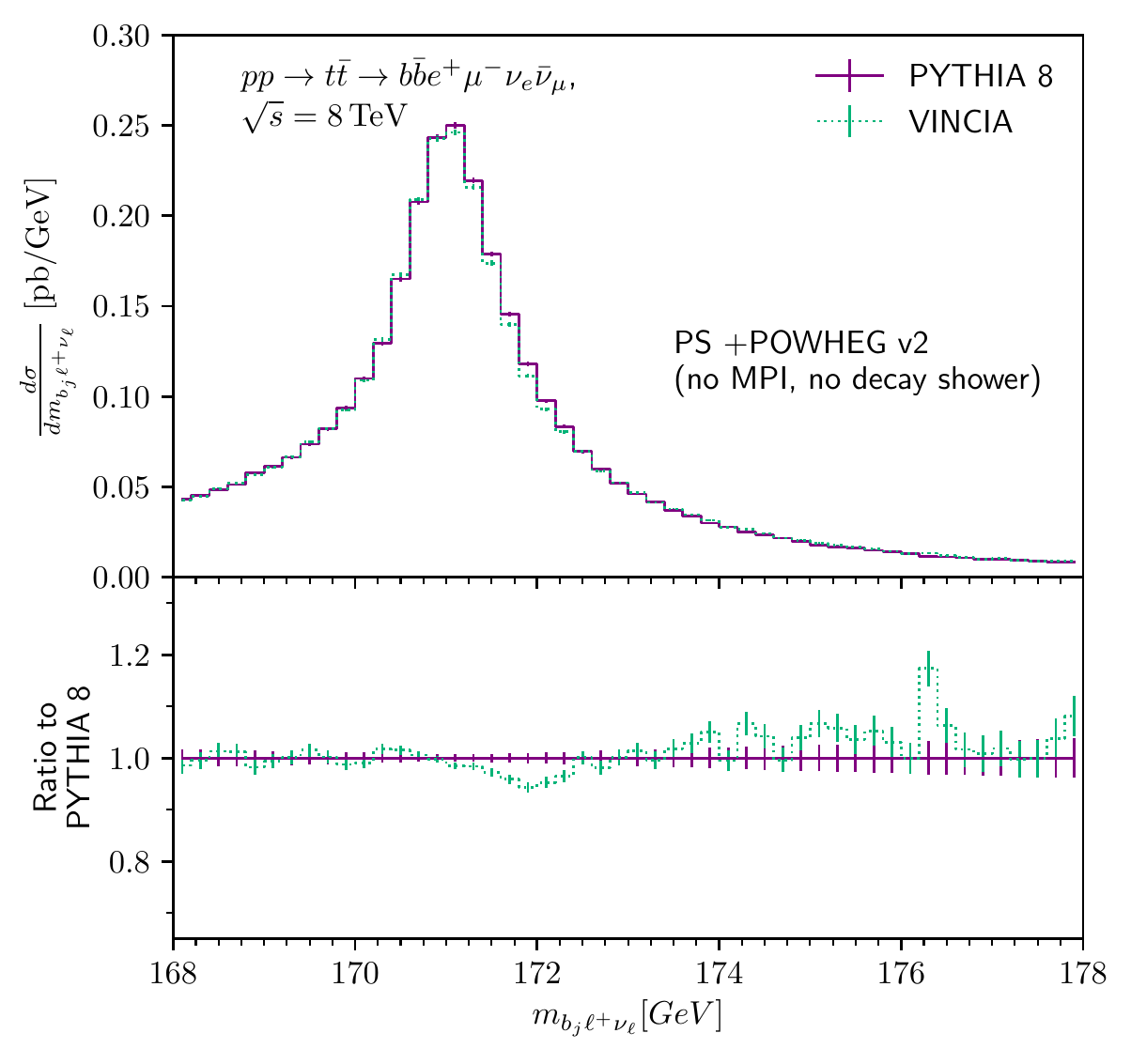}{}
  \caption{}
    \label{fig:mbmlpnu_nompi_noresshower}
 \end{subfigure}
\caption{As for \cref{fig:mblpnu_mpi} but with \subref{fig:mbmlpnu_nompi} underlying event turned off, and \subref{fig:mbmlpnu_nompi_noresshower} also with the resonance decay shower turned off.
}
\label{fig:mblpnu_ps}
\end{figure}

\begin{figure}[t]
 \centering
  \begin{subfigure}{0.45\textwidth}
 \centering
 \includegraphics[width=\textwidth]{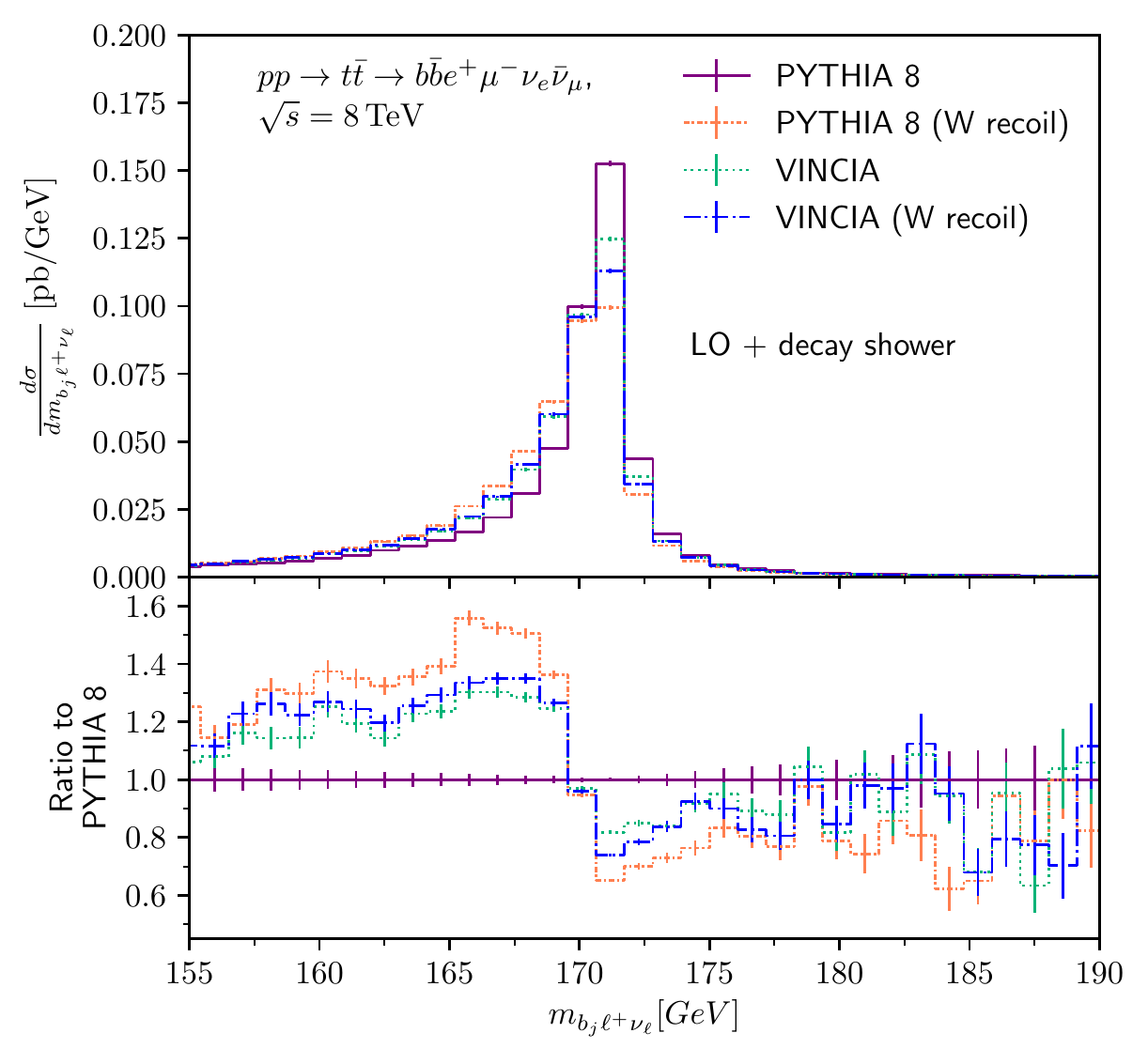}{}
  \caption{}
  \label{fig:mbmlpnu_lops}
 \end{subfigure}
  \begin{subfigure}{0.45\textwidth}
 \centering
 \includegraphics[width=\textwidth]{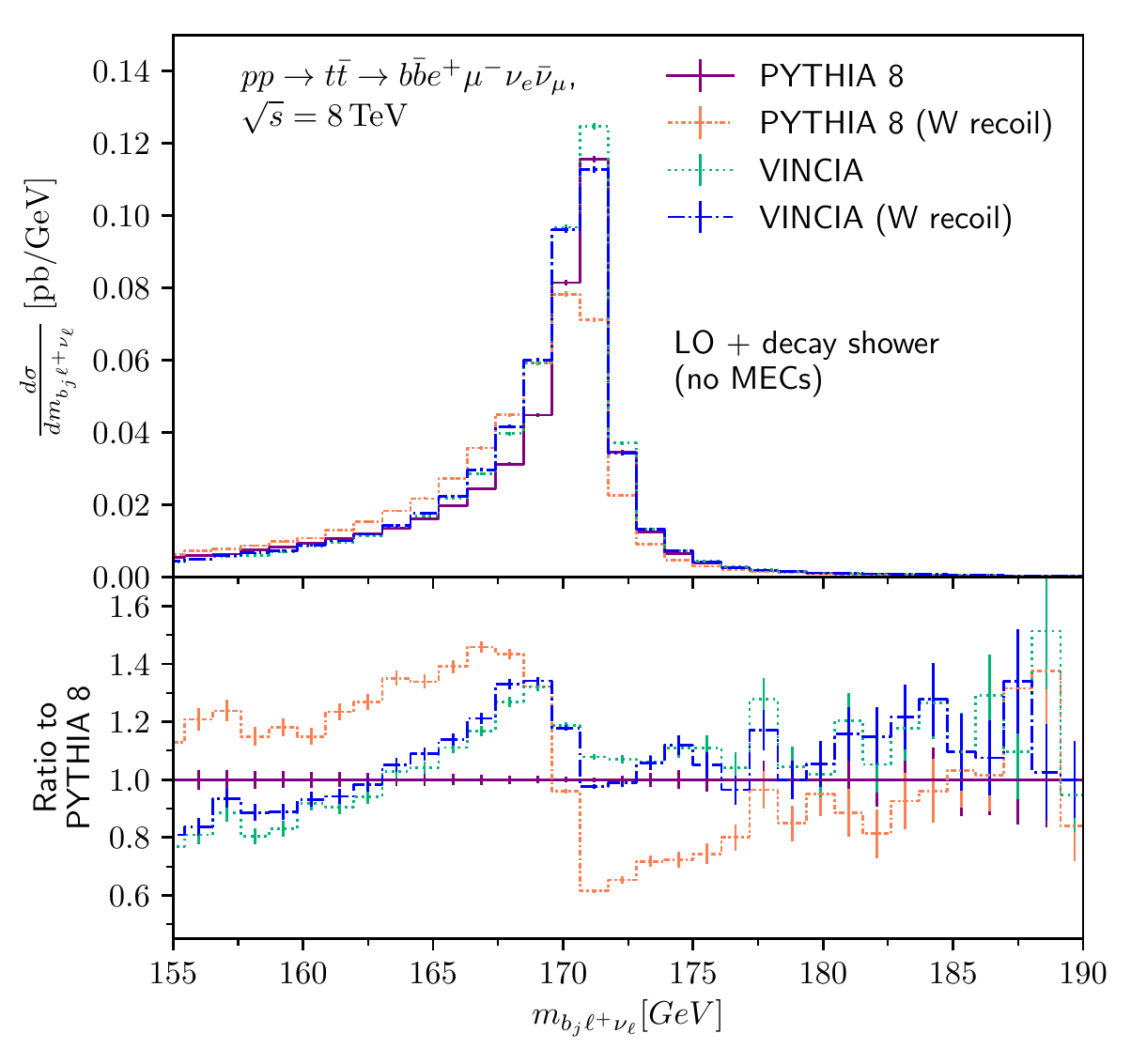}{}
  \caption{}
  \label{fig:mbmlpnu_lops_nomecs}
 \end{subfigure}
\caption{
Plots showing the leading-order differential cross section as a function of the invariant mass of the $b_j \ell^+ \nu_\ell$ system in dileptonic top pair production at the 
LHC with $\sqrt{s}=8$ TeV. The effect of the resonance decay shower is compared between \py~ and \vin~ for two choices of kinematic map. In \subref{fig:mbmlpnu_lops_nomecs}
\py's matrix-element corrections have been switched off.}
\label{fig:mblpnu_lops}
\end{figure}

\begin{figure}
\centering
\begin{subfigure}{0.45\textwidth}
 \centering
 \includegraphics[width=\textwidth]{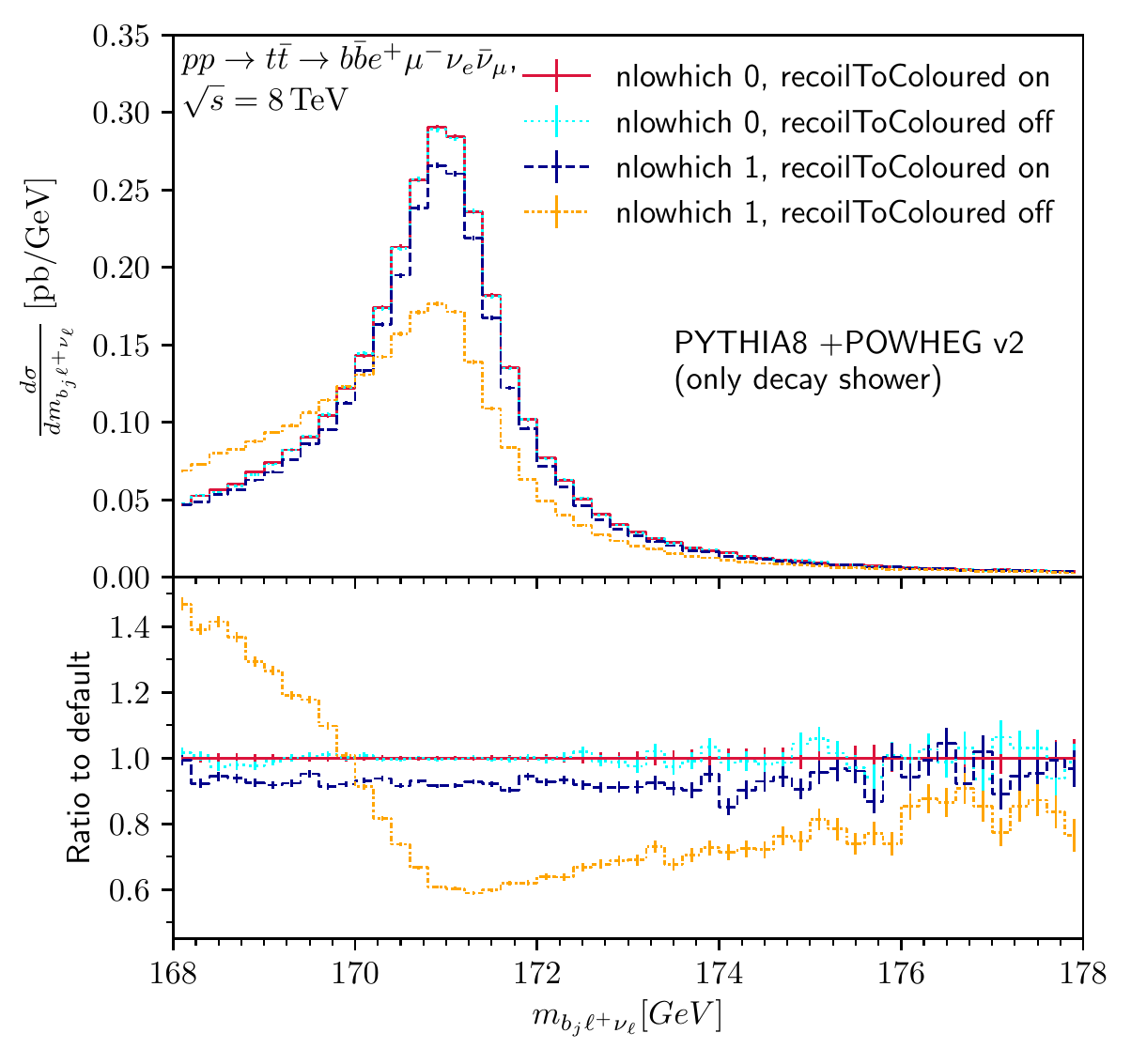}{}
  \caption{}
  \label{fig:mbmlpnu_nlowhich}
 \end{subfigure}
  \begin{subfigure}{0.45\textwidth}
 \centering
 \includegraphics[width=\textwidth]{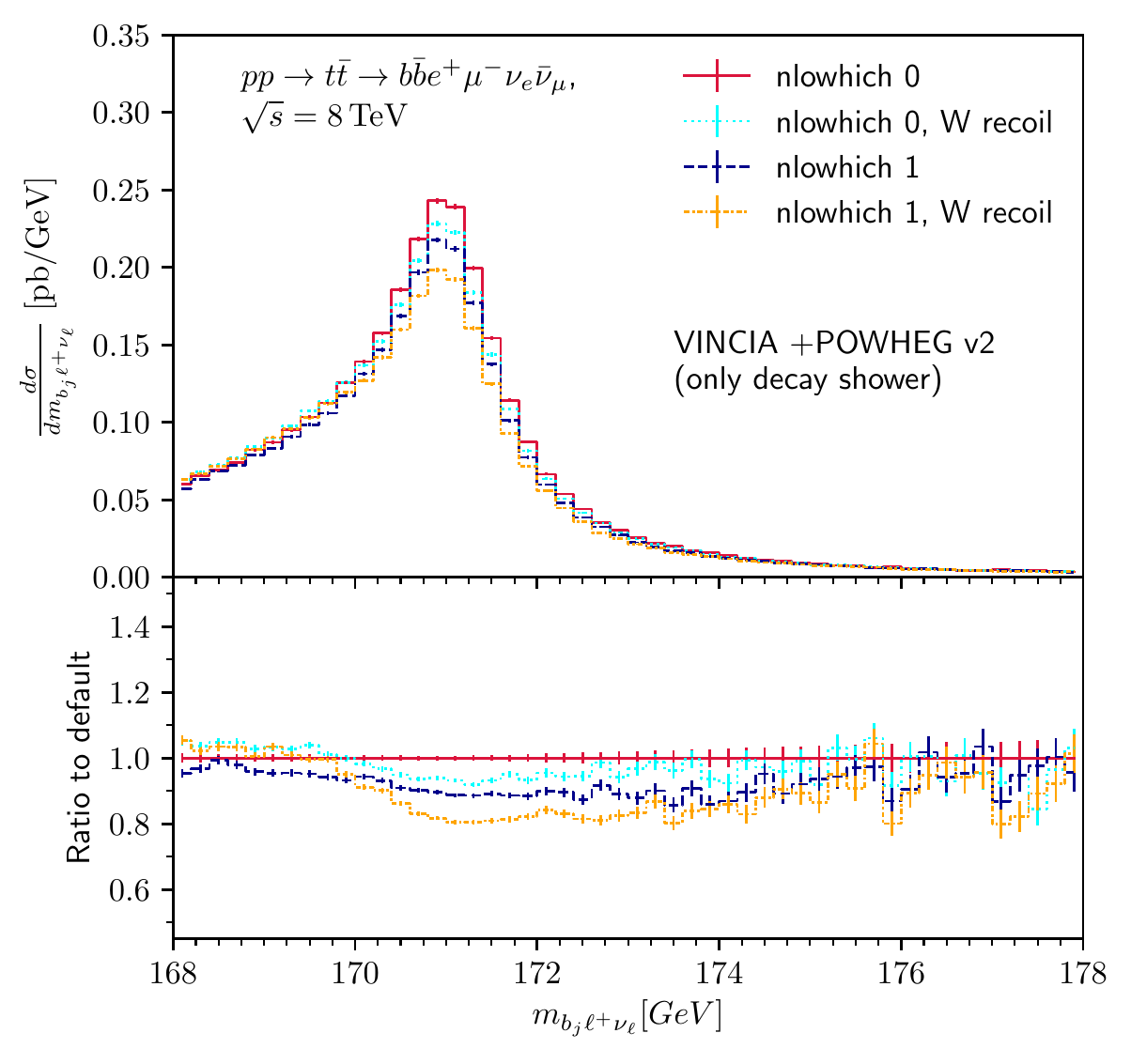}{}
  \caption{}
  \label{fig:mbmlpnu_nlowhich_vin}
 \end{subfigure}
\caption{Plots showing the differential distribution of the invariant mass of the $b_j \ell^+ \nu_\ell$ system in dileptonic top pair production at the 
LHC with $\sqrt{s}=8$ TeV, with \py~ (\subref{fig:mbmlpnu_nlowhich}) and \vin (\subref{fig:mbmlpnu_nlowhich_vin})  matched to NLO accuracy using \pwgii.
Comparisons of alternative settings of \texttt{nlowhich} combined with different choices of kinematic map are made.}
\label{fig:mblpnu_nlowhich}
\end{figure}

\begin{figure}[t]
 \centering
 \includegraphics[width=0.45\textwidth]{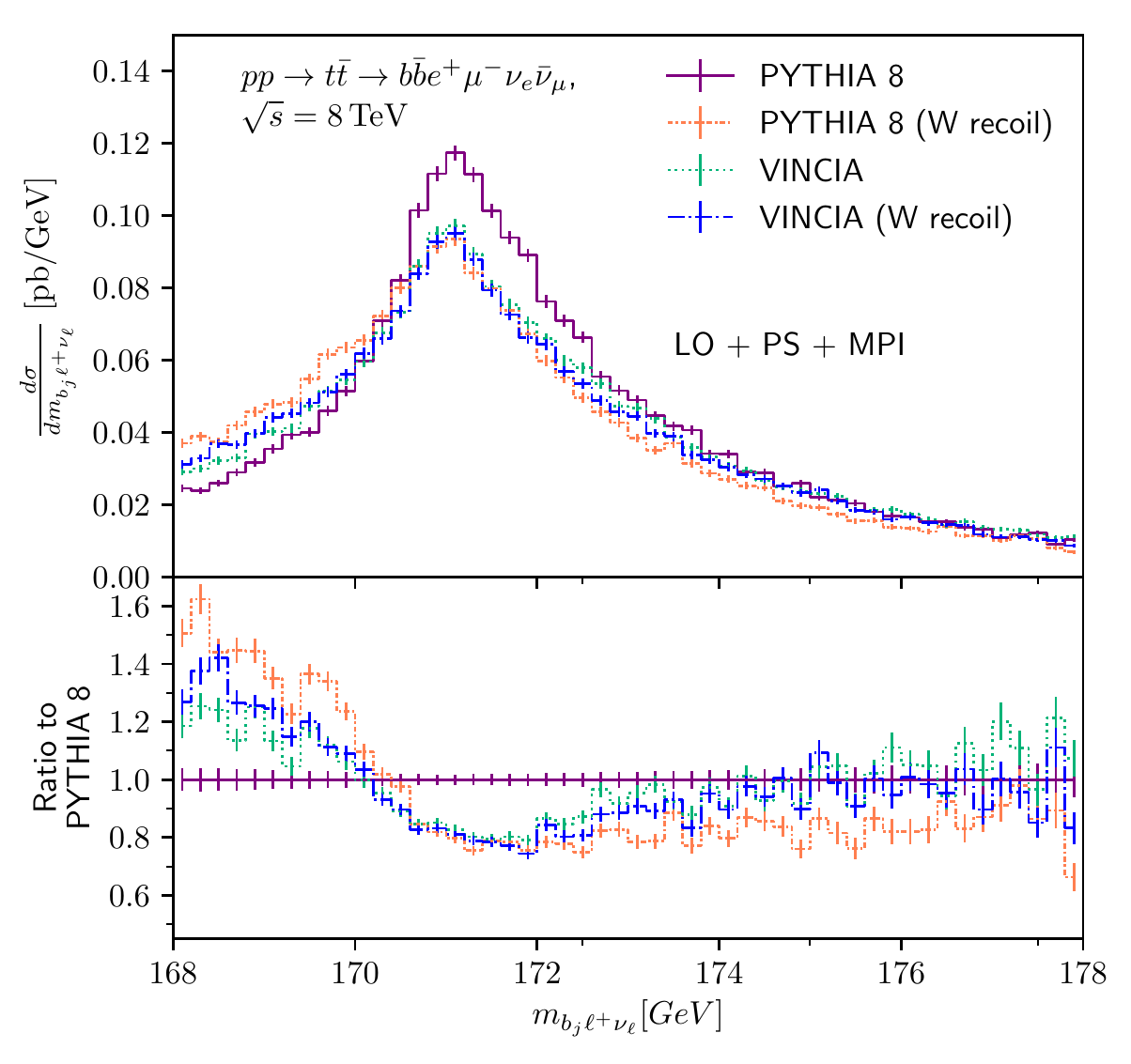}{}
\caption{As for \cref{fig:mblpnu_mpi}, but at leading order accuracy.}
\label{fig:mblpnu_lops_mpi}
\end{figure}

We proceed to investigate the effect of the parton shower in distributions that pertain to the top mass measurement. 
For \cref{fig:mblpnu_mpi,fig:mblpnu_ps,fig:mblpnu_lops,fig:mblpnu_nlowhich,fig:mblpnu_lops_mpi,fig:mblp}, the setup of our analysis is intended to be similar to that performed in \cite{Ravasio:2018lzi}, 
since it is worthwhile to reproduce the large differences observed therein. 
Specifically we consider $p\bar{p}\rightarrow t\bar{t} \rightarrow{b e^{+} \nu_e \bar{b} \mu^{-} \bar{\nu}_\mu}$ at $\sqrt{s}=8$ TeV.
We require at least two $b$-jets with ${p_T} > 30$ GeV and $|\eta|<2.5$ constructed using 
the anti-$k_T$ algorithm with $R = 0.5$. The leptons are required to have
${p_T} > 20$ GeV and $|\eta|<2.4$. In addition, the neutrino was required to have $p_T > 5$ GeV
and $|\eta|<2.4$ (relative to \cite{Ravasio:2018lzi} where no cut was placed on the neutrino).
Again, the analysis is performed at the Monte Carlo ``truth''-level, using only the correct pairing of the lepton and $b$-jet and assuming we can perfectly reconstruct the neutrinos' momenta. 

Using $b_j$ to denote a reconstructed $b$-jet, we start by analysing the invariant mass of the top decay system composed of $b_j \ell^+ \nu_\ell$. 
In \cref{fig:mblpnu_mpi} we show the differential cross section (matched to NLO using \pwg~ as described above) at parton level, prior to hadronisation
but including underlying event. There are considerable shape differences between the three generators shown.
\herwig~ gives rise to a distribution shifted towards lower masses, while \py~8 is shifted towards higher masses. 
\vin~ is hybrid between the two, giving an overall broader spectrum. Given the significant differences that arise here, we now make some concerted effort
to disentangle the different driving forces of these shape differences.

As a start, it is possible to isolate the primary differences as arising from two different sources, namely the resonance decay shower, and from underlying event,
as we demonstrate in \cref{fig:mblpnu_ps}. First removing the underlying event in \cref{fig:mbmlpnu_nompi} we find that qualitatively \vin~ and \herwig~ give
similar distributions relative to \py~8, although \herwig~ predicts a somewhat softer spectrum; however all converge towards larger invariant masses.
When in addition we compare \vin~ and \py~ with the parton shower turned off in the decay of the resonances as shown in \cref{fig:mbmlpnu_nompi_noresshower}, we observe that \vin~ and \py~
are now in strong agreement. 
We conclude that while differences in MPI modelling are largely responsible for driving differences at larger invariant masses, 
differences towards lower invariant masses arise from  the resonance decay.
The latter occurs due to differing amounts of out-of-cone radiation from the $b$-jet, as we now examine in further detail.

Turning off both initial- and final-state radiation in production (as well as underlying event) to focus on the resonance-decay shower,
we consider the impact of the recoil strategy employed, starting at leading-order accuracy. We compare the default options for \py~ and \vin~
to the option where the $W$ boson takes all the recoil from every emission (as described in \cref{sec:validate_kinmap}) in \cref{fig:mbmlpnu_lops}.
(We also show a larger range on the $x$-axis to make the effects easier to see.) When placed on an equal footing in this manner, we find that \py~ and \vin~ perform similarly - if anything 
\py~ now has a slightly broader spectrum, shifted to lower invariant masses. 

The effect of the recoil strategy on \py~ is fairly dramatic. We interpret this as being due to the phase space available for branching being limited by the invariant mass of the dipole, in which the choice of recoiler plays a vital role. The $W$ is anticollinear to the dominant direction for radiation and hence offers a relatively large phase space, in particular for wide-angle radiation, while coloured partons (as in the default choice of recoiler) tend to be more collinear and hence have smaller phase spaces from the second emission onwards. 
Thus by default, \py~ has a lower capacity to produce the kind of hard, out-of-cone radiation that has the potential to reduce the reconstructed invariant mass
(even if the branchings that do occur result in slightly broader jets).
By comparison, \vin's two recoil strategies perform similarly, because even in the default option the phase space for the RF antenna after the first emission is still set by the ``crossed top'' system which contains the $W$, and the $W$ continues to take some of the recoil.

These differences become even more pronounced when \py's matrix-element corrections are switched off, as shown in \cref{fig:mbmlpnu_lops_nomecs}
This is the consistent with the finding 
of \cref{sec:bjetprofiles} that matrix-element corrections are effectively correcting for coherence and reduce the amount of out-of-cone radiation.
We conclude that the region of low invariant mass is driven by a combination of the recoil strategy, and formally subleading corrections in the splitting kernels.

We find that MECs primarily influence the first branching, and we see no effect from modifying \texttt{TimeShower:MEafterFirst} to only turn off corrections after the first emission. 
On the other hand, alternative recoil strategies only affect secondary (or later) emissions, and thus we expect the latter to persist when matching to NLO, as we now investigate.

We compare both the default option where \pwg~may generate the hardest emission in decay, to the case where this behaviour is turned off entirely by using:
\begin{lstlisting}
nlowhich = 1
\end{lstlisting}
In the latter case, the parton shower is always responsible for generating the hardest emission in decay, and otherwise the two are identical.
However, when MECs are applied in \py~ the only difference between the shower and \pwg~ should be virtual corrections, which should not significantly affect the shape of
the distribution. The naive expectation then is there should be little effect from modifying \texttt{nlowhich}. In fact, the contrary is true, as demonstrated
for \py~8 in \cref{fig:mbmlpnu_nlowhich}.

In going from \texttt{nlowhich = 1} to \texttt{nlowhich = 0} for the default kinematic map, there is a change in normalisation which is consistent with 
adding virtual corrections. However, while for \texttt{nlowhich = 1} there remains a significant effect from activating the alternative recoil strategy 
by switching \texttt{TimeShower:recoilToColoured  = off}, this flag has no effect for \texttt{nlowhich = 0}.

For \vin, on the other hand, the effect of employing the $W$ recoil strategy is consistent with the picture at leading order, regardless of whether \pwg~ corrects the first emission in decay, as shown in \cref{fig:mbmlpnu_nlowhich_vin}.

This surprising observation has an explanation in how \py~ interprets the \texttt{recoilToColoured} flag.
When the dipoles in the resonance decay system are set up prior to the commencement of the shower, if there exists any unconnected colour tag for a parton $i$ in the
final state, a recoiler $j$ is simply selected from the available set of final state particles, by minimising the invariant:
\begin{equation}
 (p_i + p_j)^2 - m_j^2 = 2 p_i p_j.
\end{equation}
It is only \textit{after} \py~ has performed an emission that the \texttt{recoilToColoured} flag is inspected.
If at this stage the current recoiler is uncoloured, if \texttt{recoilToColoured = on} then only coloured recoilers are considered in a first step and uncoloured ones only allowed if no coloured ones are available. 

For internal events, or when \texttt{nlowhich = 1}, the system is simply $\{b,W\}$, and the $W$ must be selected as the recoiler for the dipole involving the $b$ quark.
It is only for secondary emissions that \texttt{recoilToColoured} may have an impact (as also noted in \cref{sec:validate_kinmap}).

However,  when \texttt{nlowhich = 0} and the system now contains an additional gluon, there is an ambiguity for the dipole between this
gluon and the resonance in which recoiler to select. In the majority of events, the above invariant is minimal for the $b$ quark rather than for the $W$ boson (since the gluon tends to be more collinear with the $b$ than with the $W$), and the former is therefore selected as the recoiler.
Furthermore, once  the $b$-quark has been selected as the recoiler, it is impossible for \texttt{recoilToColoured} to have any effect for subsequent emissions. 
This explains why the corresponding results in \cref{fig:mbmlpnu_nlowhich} for \texttt{nlowhich = 0} are identical: \py~treats both the same. 

Physically the impact of selecting the $b$-quark in place of the $W$ boson, is as follows. Since the gluon tends to be more collinear with the $b$ than with the $W$,
the former choice results in a smaller phase space for radiation, and produces much less out-of-cone radiation than the latter. Thus the former results in a narrow invariant mass 
distribution, and this is precisely what is observed in \cref{fig:mbmlpnu_nlowhich}.

Therefore, it is not that there is no effect from varying the recoil strategy for the resonance decay shower in \py~when \texttt{nlowhich = 0}, but rather
that at present there is no mechanism by which such a variation may be performed. We plan to implement such an option in \py~in a follow-up to this work. 
 
To conclude our discussion and make contact with the distribution in  \cref{fig:mblpnu_mpi} we  compare \py~ and \vin~ at leading order for different choices of kinematic map, now with the full shower
and including MPI, in \cref{fig:mblpnu_lops_mpi}. The kinematic map still has an impact towards lower invariant masses,
but MPI dominates towards larger invariant masses: the results converge for all choices of map and shower (albeit slightly more slowly for the $W$ recoil map in \py).
For the default choices of map, the relative size of differences between \vin~ and \py~ is fairly similar in LO and at NLO. 

We note that for \herwig~7 the broadening effect from MPI is slightly smaller than in \py~ and \vin~ (which we see by comparing \cref{fig:mblpnu_mpi,fig:mbmlpnu_nompi}. We deem it beyond the scope of this paper to study MPI effects in detail but note that an eventual follow-up study could well include in-situ measurements of the underlying event in top events such as the one by CMS~\cite{Sirunyan:2018avv}. 

Finally we comment upon the bump in the peak region that is only present for \herwig~7, that has also been observed elsewhere \cite{Ravasio:2018lzi,FerrarioRavasio:2019vmq}.
It was recently noted  \cite{FerrarioRavasio:2019vmq} that this bump is not present for \herwig~6.5 \cite{Corcella:2000bw}; the authors  of \cite{FerrarioRavasio:2019vmq} ascribe it to differences in
the ordering variable between the two versions, and a potential cutoff mismatch between the shower and \pwg. We use the same matching settings
as in \cite{Ravasio:2018lzi} so we would be afflicted by the same mismatch.

\subsection{A more realistic analysis}
\label{fig:realistic}

In the previous section we discussed in detail the consequences of alternative parton showers for the differential distribution of the 
invariant mass of the $b$-jet, charged lepton and neutrino system, $m_{b_j \ell \nu_\ell}$, in the dilepton channel for $t \bar{t}$ production. 
The dilepton channel is a particularly clean arena in which to perform a measurement of the top quark, primarily because
the charged leptons carry information about the top quark kinematics without suffering from hadronic uncertainties (such as the jet energy scale).
However in dilepton production, in practice we cannot reconstruct the momenta of the neutrinos. 
Thus in direct measurements of the top quark mass it is standard practice in both CMS \cite{Chatrchyan:2013boa,Sirunyan:2017idq} and ATLAS \cite{Aad:2015nba,Aaboud:2016igd}
to instead measure the invariant mass of the $b$-jet and charged lepton, $m_{b_j \ell}$, and extract the top quark mass 
by performing a fit of shower Monte Carlo event generators.
In particular, the distribution of $m_{b_j \ell}$ exhibits a kinematic endpoint (to which it falls sharply) that is sensitive to the value of the top quark mass.
Thus in order to determine the impact of differences in physics modelling 
between different generators we now consider this observable, and examine the sensitivity of the endpoint.

\begin{figure}[t]
 \centering
 \begin{subfigure}{0.45\textwidth}
 \centering
 \includegraphics[width=\textwidth]{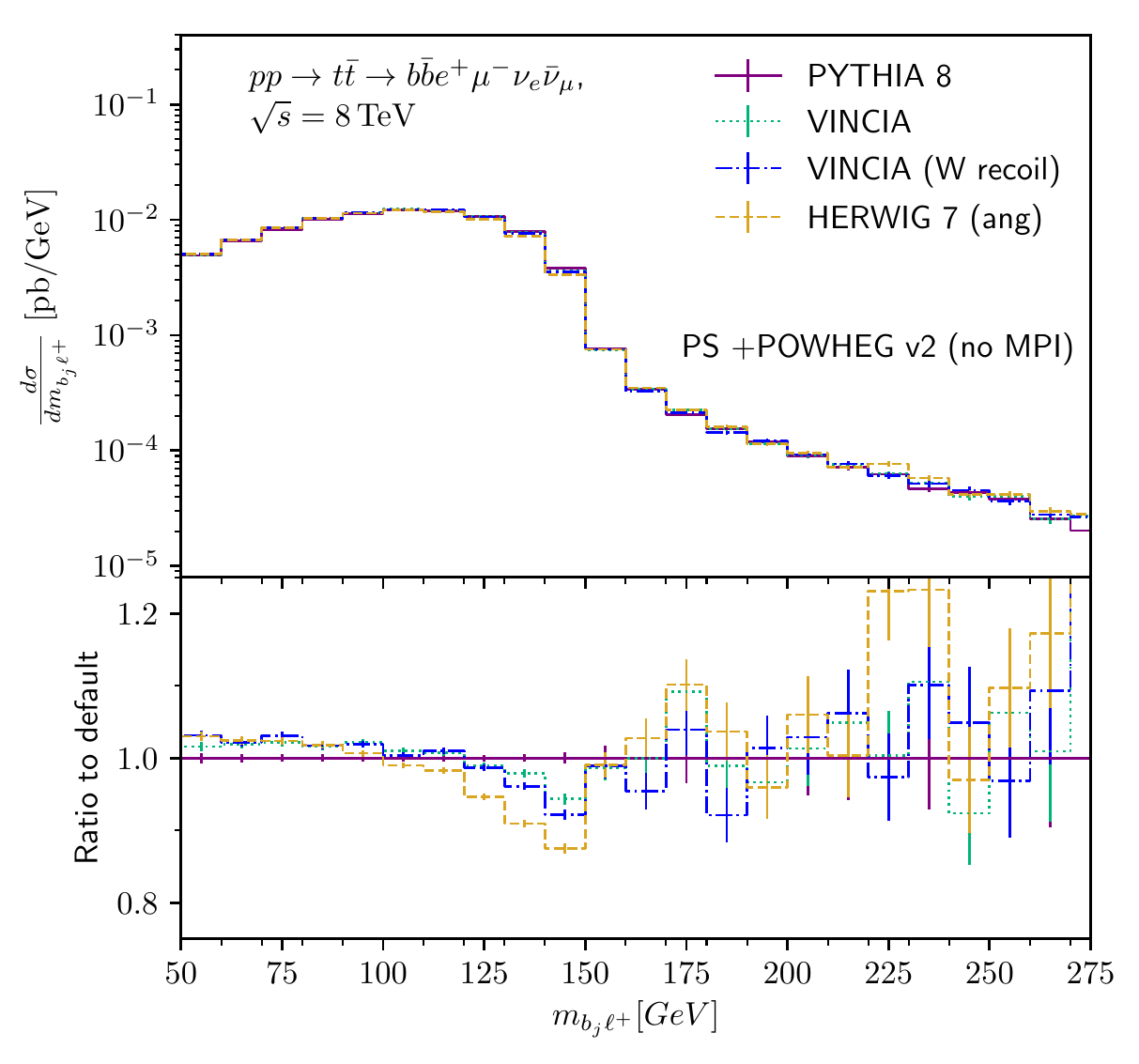}{}
  \caption{}
  \label{fig:mblp_nompi}
 \end{subfigure}
 \begin{subfigure}{0.45\textwidth}
 \centering
 \includegraphics[width=\textwidth]{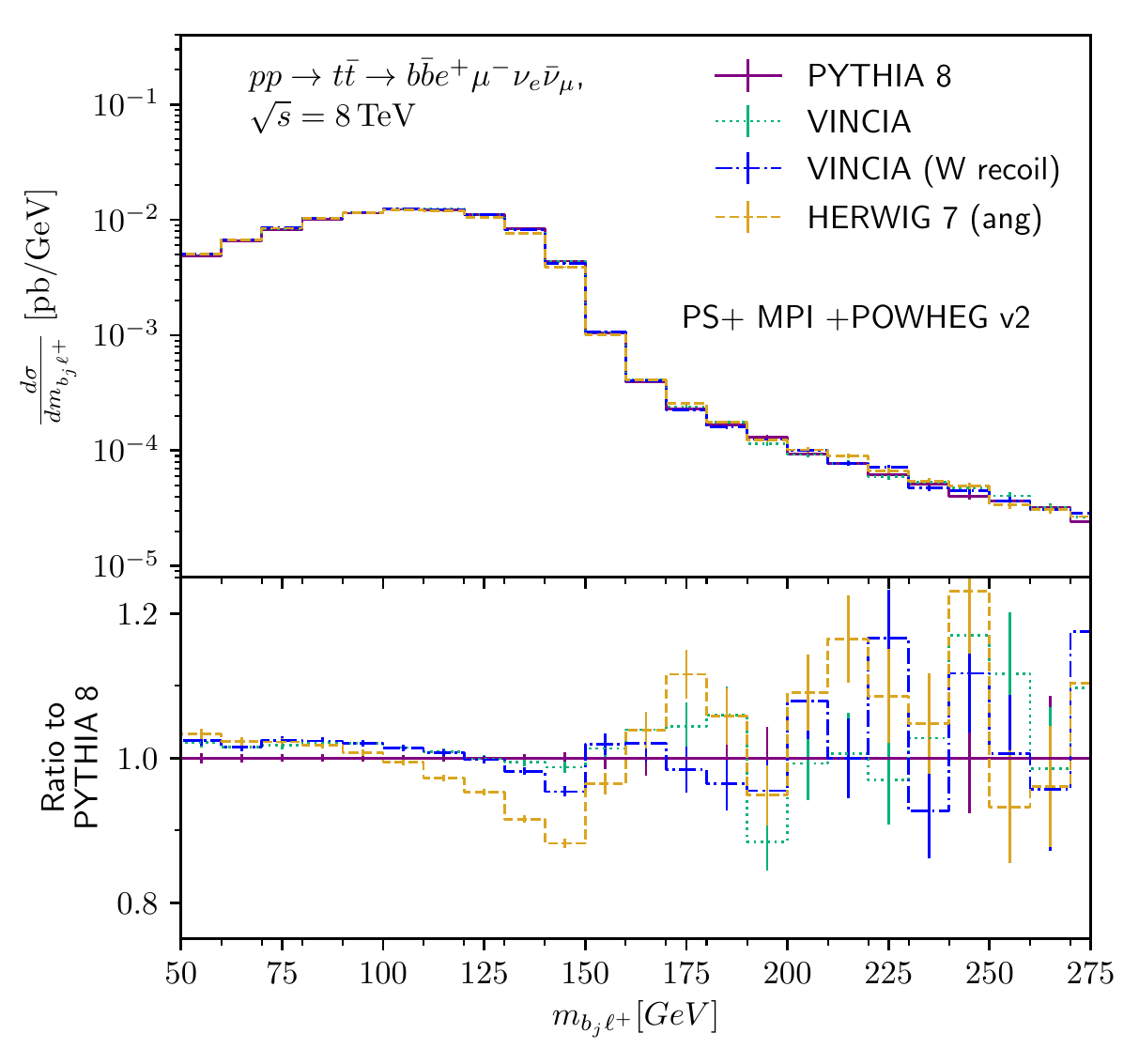}{}
  \caption{}
    \label{fig:mblp_mpi}
 \end{subfigure}
\caption{Plots showing the differential cross section as a function of the invariant mass of the $b_j \ell^+$ system,
without (\subref{fig:mblp_nompi}) and with (\subref{fig:mblp_mpi}) MPI.
The setup is the same as \cref{fig:mblpnu_mpi}.}
\label{fig:mblp}
\end{figure} 

We start by considering this observable at parton-level, namely prior to hadronisation, using the same analysis setup as in the previous section.
At this stage we still perform a ``truth''-level analysis,  identifying the ``correct'' pairings of the $b$-jet and the lepton based on their respective
charges. The differential cross section in the invariant mass of the $b_j \ell^+$ system is shown in \cref{fig:mblp}
without (\subref{fig:mblp_nompi}) and with (\subref{fig:mblp_mpi}) MPI. 

In the former, we  observe  that the sensitivity of the low-mass region to the kinematic map has been greatly reduced, to the level of a few percent, as may be seen by comparing the
two options available for \vin. The primary location that is sensitive to the kinematic map is the endpoint itself: \vin~ falls off more quickly than \py.
An effect that is qualitatively similar, although larger at the 10\% level, is the difference of \herwig~ with respect to \py~ (the former also falling off more quickly).
This is consistent with the observation that the mass peak for \herwig~ in \cref{fig:mbmlpnu_nompi} also falls off more quickly.

After the inclusion of MPI, the relative difference induced by changing the kinematic map is reduced, while the difference with respect to \herwig~ persists. This is consistent
with the picture seen in \cref{fig:mblpnu_mpi}, where \vin~ and \py~ converge in the high-invariant-mass region, while \herwig~ remained qualitatively different.
This perhaps implies that the modelling of MPI is the dominant uncertainty in the location of the endpoint.
However we emphasise that it is difficult to disentangle the two physics effects since the sensitivity to both has essentially been ``squeezed'' into a single kinematic region.
We therefore repeat that dedicated studies of the underlying event in top-pair events, such as~\cite{Sirunyan:2018avv},
may be relevant to constrain the ambiguity associated with the MPI component.

\begin{figure}
 \centering
 \includegraphics[width=0.45\textwidth]{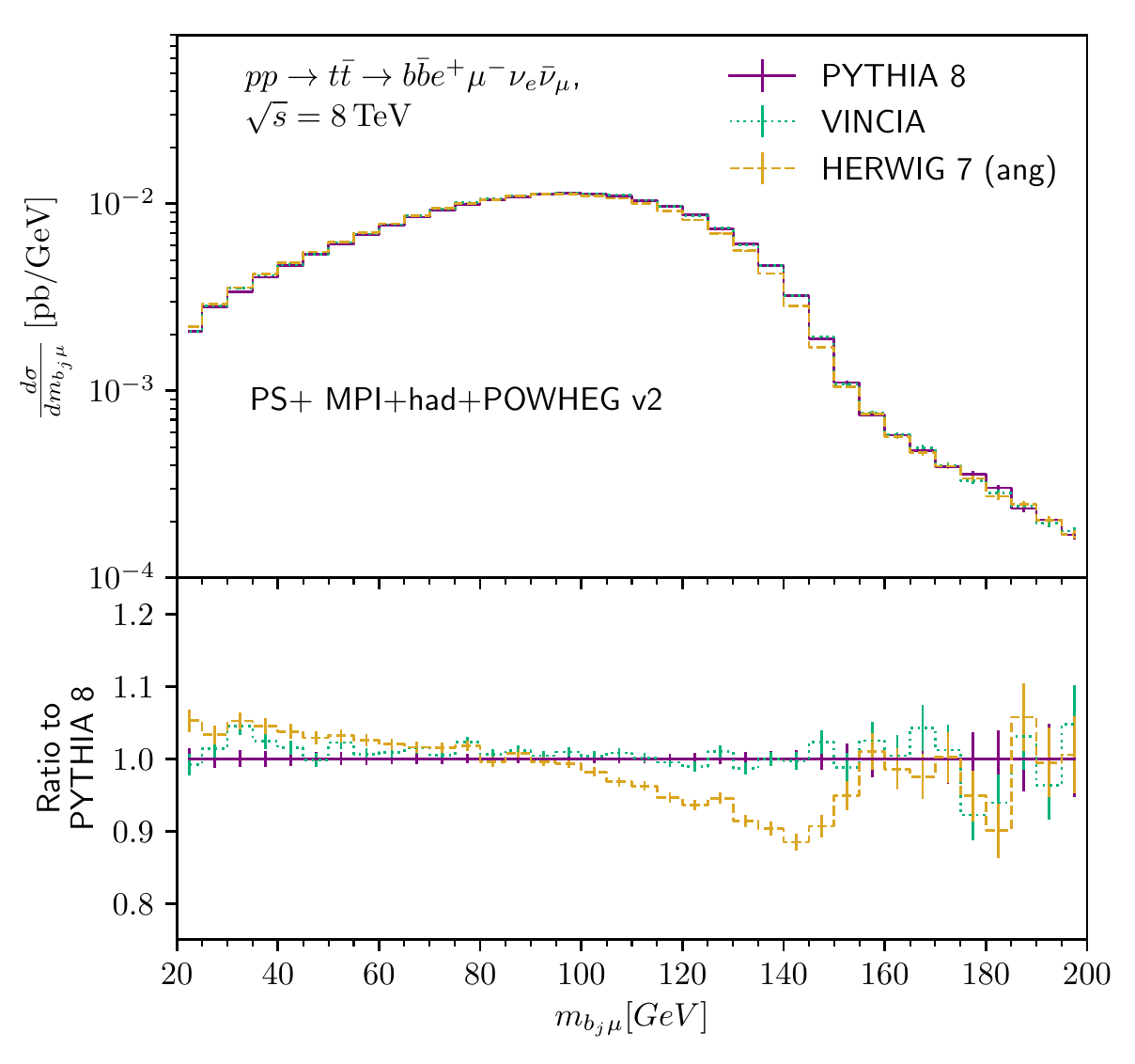}
  \caption{Plot showing the differential cross section as a function of the invariant mass of the $b_j \mu$ system,
  in dileptonic top pair production at the LHC with $\sqrt{s}=8$ TeV, with the parton shower matched
  to NLO accuracy. Results are shown at particle level.}
  \label{fig:mbjmu}
\end{figure}

We now proceed to perform a particle-level analysis. Although the setup is similar to that of  \cref{sec:psnlo}
we deem it inappropriate to overly interpret results based on perfect reconstruction from the event record.
In particular, we no longer assume we can find the correct pairings of the $b$-jet and charged lepton. 
Instead, the invariant mass for each possible $b$-jet-lepton pairing (from the hardest two  of each) is calculated, and
the set of pairings for which the average invariant mass is minimal is chosen
(this is the method used in \cite{Aaboud:2016igd}).

The cut-selection used was chosen to be similar to that used in \cite{Aaboud:2017ujq},
and the analysis was implemented in \rivet.
\footnote{Furthermore, we made use of the corresponding public analysis \texttt{ATLAS\_2017\_I1626105}
where possible to keep the implementation as similar as possible.}
The event was required to have two $b$-jets and two opposite-sign charged leptons with different flavours.
The $b$-jets were constructed with the anti-$k_T$ algorithm with $R=0.4$, and were required to have ${p_T} > 25$ GeV and $|\eta|<2.5$.
The charged leptons were dressed with any radiation from photons with a radius of $\Delta R=0.1$.
Both charged leptons were required to have ${p_T} > 25$ GeV and $|\eta|<2.5$.
Jets were vetoed if there was a charged lepton within a radius of $\Delta R = 0.2$, and leptons within 
a radius of $\Delta R = 0.4$ from an accepted jet were vetoed.

In \cref{fig:mbjmu} we show the lepton-jet invariant mass for the pairing that includes the muon. 
Qualitatively, the results are fairly similar to \cref{fig:mblp_mpi}, however the entire distribution is shifted
down in mass, and the spectrum is broader. It is therefore not surprising that the region over which \herwig~ exhibits
differences with respect to \py~ is also broader, although reaching a similar maximal relative difference of about 10\%.
The largest differences remain in the region of the endpoint, with the low invariant mass region continuing to exhibit relatively little sensitivity
to the different shower models. We emphasise that this is precisely the region that is fitted to extract a measurement of the top quark mass, 
and is therefore relevant for theoretical uncertainties.

Finally, we also considered the ``stransverse mass'' variable, $m_{T2,bb}$ , originally defined in \cite{Lester:1999tx} and calculated using in-built functions
in \rivet \cite{Cheng:2008hk,Lester:2014yga}, that has been used
in direct top mass measurements by CMS \cite{Sirunyan:2017idq}.
We see a similar level of difference to \cref{fig:mbjmu}, so we do not consider it enlightening to reproduce here. 

%

\section{Summary and Outlook}
\label{sec:conclusions}

We have implemented resonance decays for \vin, an antenna-shower plugin to \py~8. 
Like traditional angular-ordered showers, the antenna-shower formalism has coherence 
built in as a fundamental tenet (even without azimuthal averaging), but does not suffer from dead zones arising
from approximate phase-space factorisations. Unlike the dipole-shower formalism 
where the soft limits are partitioned across two radiators, we can utilise the positive-definiteness
of the massive eikonal and construct our antenna functions such that they are positive-definite everywhere.

Based on arguments stemming from the antenna factorisation, we argue for a more democratic treatment of recoils from branchings
in resonance decays, namely that recoils are shared among all final-state particles in the decay system.
In addition we have implemented an alternative, but less theoretically sound, recoil strategy to allow for closer comparisons
with \py~8  and \herwig~7, in which the original uncoloured child in the decay system continue to receive all recoil.

We have used our formalism to help disentangle the causes of significant shape differences observed between generators for the reconstructed invariant mass spectrum of the top quark.
Although coherence plays a role, we find that matrix-elements corrections are essentially sufficient to restore these effects. 
We find that the differences are primarily driven by (a) the choice of recoil strategy, and (b) underlying event. Since both of these effects 
arise from ambiguities that are purely subleading, we regard the differences
as indeed representative of the theoretical uncertainty. Our recommendation therefore is that variations of these aspects of event generation should
be performed in order to obtain trustworthy uncertainties, in particular where comparisons to event generators are used to extract measurements. We presume this
should have some impact upon the uncertainty on the top quark mass measurement (although we do not attempt to estimate this here).

We comment here that our results may be dependent upon the exact choice of radius used in the definition of the $b$-jet. 
Increasing the jet radius may decrease the impact of wide-angle radiation, but on the other hand may increase contamination from MPI.
Investigating this dependence was beyond the scope of this work, however it is likely that such a study could prove worthwhile.
Additionally, nowhere did we consider the impact of spin correlations. 
This is a worthwhile topic for consideration in its own right.
Finally, we note that while the focus of this paper has been on QCD radiation, our treatment has been combined 
with recent work by Kleiss and Verheyen on antenna-based multipole QED showers~\cite{Kleiss:2017iir},
so that \vin~ includes both QCD and (fully coherent) QED shower branchings within a single interleaved framework. 

It should be clear that further developments to parton showers are required, 
since it is at this stage of generation at which uncertainties arise.
In the context of resonance decays, further developments of \vin~ are underway on matrix-element corrections, sectorised showers, electroweak corrections, and finite-width effects (to account for the interference between
production and decay). The main long-term goal for us remains  improvements to the perturbative accuracy of the parton shower itself.

\subsection*{Acknowledgements}
We thank Silvia Ferrario-Ravasio for advice on using \pwgii~; we also thank Silvia, Johannes Bellm, and Peter Richardson
for providing an interface for using \herwig~7 with \pwgii~ and for advice on \herwig~7 settings.
Thanks to Malin Sj\"{o}dahl, Tobj\"{o}rn Sj\"{o}strand and Simone Amoroso for helpful comments while the manuscript was being completed, and thanks to Stefan H\"{o}che for providing clarification on the 
treatment of resonance decays in \sherpa.
HB is funded by the Australian Research Council via Discovery Project DP170100708 -- ``Emergent Phenomena in Quantum Chromodynamics''. PS is supported in part by the Australian Research Council, contract FT130100744.
This work was also supported in part by the European Union's Horizon 2020 research and innovation programme under the Marie Sklodowska-Curie grant agreement No 722105 -- “MCnetITN3”.

\appendix

\section{Massive Helicity-Dependent Initial-Final Antenna Functions}
\label{app:antennae}

For the sake of completeness, here we detail all massive initial-final antenna functions which have been changed relative to \cite{Fischer:2016vfv,Fischer:2017htu}.
Aside from the addition of mass effects (obtained from crossing symmetry of massive final-final antennae \cite{GehrmannDeRidder:2011dm}),
some finite terms have been added in order to ensure all individual helicity antenna functions are positive-definite everywhere.

For compactness of notation, we define the dimensionless helicity antenna function:
\begin{equation}
 \tilde{a}_{h_A h_K, h_a h_j h_k} \equiv s_{AK}\, a( h_A h_K \to h_a h_j h_k)
\end{equation}

\subsection{QQemitIF}

The helicity-averaged antenna function for $q_A q_K \rightarrow q_a g_j q_k$ is: 

\begin{align} 
  a= &  
  \frac{1}{s_{AK}}\left[\frac{(1-y_{aj})^2 + (1-y_{jk})^2}{y_{aj} y_{jk}}  \right . \nonumber \\
   & \left. \qquad-\frac{2\mu^2_a}{y_{aj}^2}\left((1-y_{jk})\left(1- \frac{y_{aj}}{2}\right) -\frac{y_{aj}}{2}(1-y_{aj})  \right) \right. \nonumber \\
    &\left. \qquad - \frac{2\mu_k^2}{y_{jk}^2}\left(1 - \frac{y_{jk}}{4}(2-y_{jk})\left(2+\frac{y^2_{aj}}{1-y_{aj}}\right)\right) \right. \nonumber \\
  & \left. \qquad+\frac{1}{2}(2-y_{aj})(2-y_{jk}) \right]~. 
\label{eq:antrf_qq}
\end{align}  

The individual helicity contributions are: 
\small{
\begin{eqnarray}
  \tilde{a}_{++, +++} =& \frac{1}{y_{aj} y_{jk}} -\frac{\mu^2_a}{y_{aj}^2} -\frac{\mu^2_k}{(1-y_{aj})y_{jk}^2}, \\
  \tilde{a}_{++, +-+} =& \frac{(1-y_{aj})^2 + [(1-y_{jk})^2 -1](1-y_{aj})^2 }{y_{aj}y_{jk}} \nonumber \\    
&-\frac{\mu^2_a (1-y_{jk}-y_{aj})^2}{y_{aj}^2} \nonumber \\ 
&-\frac{\mu^2_k (1-y_{aj})(1-y_{jk})^2}{y_{jk}^2} ,\\
\tilde{a}_{++, --+} = & \frac{\mu_a^2 y_{jk}^2}{y_{aj}^2},\\
\tilde{a}_{++, ++-} = & \frac{\mu_k^2 y_{aj}^2}{(1-y_{aj})y_{jk}^2},\\
\tilde{a}_{+-, ++-} = & \frac{(1-y_{aj})^2}{y_{aj} y_{jk}} -\frac{\mu_a^2(1-y_{aj})}{y_{aj}^2} -\frac{\mu_k^2(1-y_{aj})}{y_{jk}^2}, \\
\tilde{a}_{+-, +--} = & \frac{(1-y_{jk})^2}{y_{aj} y_{jk}}-\frac{\mu_a^2(1-y_{jk})^2}{y_{aj}^2} -\frac{\mu_k^2(1-y_{jk})^2}{y_{jk}^2(1-y_{aj})} , \\
\tilde{a}_{+-, ---} = & \frac{\mu^2_a y_{jk}^2}{y_{aj}^2} ,\\
\tilde{a}_{+-, +-+} = & \frac{\mu^2_k y_{aj}^2}{y_{jk}^2 (1-y_{aj})}~.
\end{eqnarray}
}

\subsection{QGemitIF}

The helicity-averaged antenna function for $q_A g_K \rightarrow q_a g_j g_k$ is:
\begin{align}
  a = & \frac{1}{s_{AK}}\Bigg[\frac{(1-y_{aj})^3 + (1-y_{jk})^2}{y_{aj} y_{jk}} + (1-\alpha)\frac{1-2y_{aj}}{y_{jk}} \nonumber \\
  & \qquad - \frac{2\mu^2_a}{y_{aj}^2} \left( (1-y_{jk}) - \frac{y_{aj}}{4}\left[ 1 +(2-y_{jk}-y_{aj})^2\right]   \right) \nonumber\\
  & \qquad  + \frac32 + y_{aj} - \frac{y_{jk}}{2} - \frac{y_{aj}^2}{2} \Bigg]    	
\end{align} 

The individual helicity contributions are:
\small{
\begin{eqnarray}
\tilde{a}_{++, +++} = & \frac{1}{y_{aj} y_{jk}} + (1-\alpha)\frac{1-2y_{aj}}{y_{jk}}-\frac{\mu^2_a}{y_{aj}^2}, \\
\tilde{a}_{++, +-+} = & \frac{(1-y_{aj})^3 +  (1-y_{jk})^2 - 1}{y_{aj} y_{jk}} -\frac{\mu_a^2(1-y_{jk}-y_{aj})^2(1-y_{aj})}{y_{aj}^2}\nonumber \\ & + 3 - y_{aj}^2 \\
\tilde{a}_{++,--+}  = & \frac{\mu^2_a y_{jk}^2}{y_{aj}^2} \\
\tilde{a}_{+-, ++-} = & \frac{(1-y_{aj})^3}{y_{aj}y_{jk}} -\frac{\mu_a^2(1-y_{aj})^2}{y_{aj}^2}, \\
\tilde{a}_{+-, +--} = & \frac{(1-y_{jk})^2}{y_{aj} y_{jk}} + (1-\alpha)\frac{1-2y_{aj}}{y_{jk}} -\frac{\mu_a^2(1-y_{jk})^2}{y_{aj}^2} \nonumber \\ &+ 2y_{aj} -  y_{jk} \\
\tilde{a}_{+-, ---} = & \frac{\mu^2_a  y_{jk}^2}{y_{aj}^2} 
\end{eqnarray}
}

\subsection{GQemitIF}

The helicity-averaged antenna function for $g_A q_K \rightarrow g_a g_j q_k$ is:
\begin{align}
  a = & \frac{1}{s_{AK}}\left[\frac{(1-y_{jk})^3 + (1-y_{aj})^2}{y_{aj}y_{jk}} + \frac{1+y_{jk}^3}{y_{aj}(1-y_{jk})}\right. \nonumber \\
  &\left. \qquad - \frac{2\mu_k^2}{y_{jk}^2}\left(1 - \frac{y_{jk}}{4}(3-3y^2_{jk}+y^3_{jk})\left(2+\frac{y^2_{aj}}{1-y_{aj}}\right)\right) \right. \nonumber \\
  &\left. \qquad  +\frac{1}{2}(2-y_{aj})(3-y_{jk}+y^2_{jk}) \right].
\end{align} 

The individual helicity contributions are:
\small{
\begin{eqnarray}
\tilde{a}_{++, +++} = & \frac{1}{y_{aj} y_{jk}} + \frac{1}{y_{aj} (1 - y_{jk})} -\frac{\mu_k^2}{y_{jk}^2(1-y_{aj})}, \\
\tilde{a}_{++, +-+} = & \frac{(1-y_{aj})^2 + [(1-y_{jk})^3 -1](1-y_{aj})^2}{y_{aj}y_{jk}} \nonumber \\ & - \frac{\mu_k^2(1-y_{aj})(1-y_{jk})^3}{y_{jk}^2} ,\\
\tilde{a}_{++, --+} = & \frac{y_{jk}^3}{y_{aj} (1-y_{jk})} ~, \\
\tilde{a}_{++,++-}  = & \frac{\mu_k^2  y_{aj}^2}{y_{jk}^2(1-y_{aj})},\\
\tilde{a}_{+-, ++-} = & \frac{(1-y_{aj})^2}{y_{aj} y_{jk}} +   \frac{1}{y_{aj} (1-y_{jk})}- \frac{\mu_k^2(1-y_{aj})}{y_{jk}^2},\\
\tilde{a}_{+-, +--} = & \frac{(1-y_{jk})^3}{y_{aj} y_{jk}}  -\frac{\mu_k^2(1-y_{jk})^3}{y_{jk}^2(1-y_{aj})},\\
\tilde{a}_{+-, ---} = & \tilde{a}_{++, --+} ~,\\
\tilde{a}_{+-, +-+} = & \tilde{a}_{++, ++-} ~.
\end{eqnarray}
 }

\subsection{XGsplitIF}
The helicity-averaged antenna function for $X_A g_K \rightarrow X_a \bar{q}_j q_k$ is:
\begin{equation}
a = \frac{1}{2m^2_{jk}} \left[ y_{ak}^2 + y_{aj}^2 + \frac{2m_{j}^2}{m_{jk}^2}\right]~.
\end{equation}

The helicity contributions are:
\begin{eqnarray}
a_{X+,X-+} = & \frac{1}{2m^2_{jk}}\left[ y_{ak}^2 -\frac{m_j^2
    y_{ak}}{m^2_{jk}(1-y_{ak})} \right]~, \\
a_{X+,X+-} = & \frac{1}{2m^2_{jk}}\left[ y_{aj}^2 -\frac{m_j^2
    y_{aj}}{m^2_{jk}(1-y_{aj})} \right]~,\\
a_{X+\to X++} = & \frac{m_j^2}{2m^4_{jk}}\left[
  \frac{y_{aj}}{(1-y_{aj})} +\frac{y_{ak}}{(1-y_{ak})} + 2\right]~.
\end{eqnarray}

\section{Method for Calculating PYTHIA's Ordering Variable} 
\label{app:ptevolCalc}

In \py~ the evolution scale for final-final dipole branchings is defined as follows:
\begin{equation}
 p^2_{\perp,\mathrm{evol}} = z(1-z)Q^2.
\end{equation}
For light branchings $aR \rightarrow (a^*  r\rightarrow) b c r$ (for radiator $a$ and recoiler $ r\rightarrow  R$),  we have 
\begin{equation}
 Q^2 = m_{a^*}^2 = (p_b +p_c)^2
\end{equation}
and $z = z_\mathrm{phys}$ is the energy fraction of the daughter $b$ in the rest frame of the $a-R$ dipole system, that may be calculated as:
\begin{equation}
 z_\mathrm{phys} = \frac{p_b\cdot (p_b+p_c+p_r)}{(p_b+p_b)\cdot (p_b+p_c+p_r)} 
 \label{eq:energyfrac}
\end{equation}

For branchings involving massive quarks $q$ (having mass $m_q$) these variables are slightly modified.
For gluon emission off a massive quark, one calculates $z_\mathrm{phys}$ according to \cref{eq:energyfrac}, and define:
\begin{equation}
  z = 1 - \frac{1-z_\mathrm{phys}}{\beta} 
\end{equation}
and
\begin{equation}
 \beta = 1 - \frac{m_b^2}{m_{a^*}^2}
\end{equation}
Further $Q^2$ is defined as
\begin{equation}
 Q^2 = m_{a^*}^2 - m_{a}^2= (p_b +p_c)^2 - p_a^2.
\end{equation}
Note that this is nearly - but not quite - the virtuality of $a^*$ because $m_{a}$ is not required to be the on-shell mass of particle $a$. In particular, this is true for resonances
such as top quarks. These variables result from the choice of generating $p_c$ according the massless kinematics, and scaling this by $\beta$.

For gluon to massive quark $g \rightarrow q\bar{q}$ splittings $Q^2$ is unchanged relative to the massless case, but for the splitting variable we have:
\begin{equation}
 z = \frac{1}{2}\left( 1 +   \frac{1}{\tilde{\beta}}(2 z_\mathrm{phys} -1) \right)
\end{equation}
where 
\begin{equation}
 \tilde{\beta} = \sqrt{1 - \frac{4m_q^2}{m_{a^*}^2}}
\end{equation}
These variables result from the choice of generating the transverse momentum of the quark, antiquark pair according the massless kinematics, and scaling by $\tilde{\beta}$.

To calculate the above in general for \vin~ requires some method of assigning the roles of radiator and recoiler. One way to do this is based on the invariants, however
in the case of emission from resonance-final antennae, we always interpret the final state particle as the radiator, since this is closest to what \py~ does. 
In addition, regardless of whether the recoil is given to a single particle or shared between several, we use the collective recoil for evaluating the above expressions.

\bibliographystyle{apsrev4-2}
\bibliography{ms}

\end{document}